\documentclass[aps,prd,preprintnumbers,groupedaddress,nofootinbib,amssymb,eqsecnum,notitlepage]{revtex4-2}
\usepackage{here}
\usepackage{array,multirow,graphicx}
\graphicspath{{figures/}}
\usepackage{amsmath,amsthm,amssymb}
\usepackage{bm}
\usepackage[dvipsnames]{xcolor}
\usepackage[normalem]{ulem}
\usepackage{comment}

\usepackage{amsfonts}
\usepackage{dcolumn}
\usepackage{stackengine}
\usepackage{braket}
\usepackage{ulem}
\usepackage[unicode=true,
 bookmarks=true,bookmarksnumbered=false,bookmarksopen=false,
 breaklinks=false,pdfborder={0 0 1},backref=false,colorlinks=true]
 {hyperref}
\definecolor{urlblue}{rgb}{0,0,0.9}\definecolor{linkblue}{rgb}{0,0,.8}\definecolor{linkgreen}{rgb}{0,0.45,0}\definecolor{linkpurple}{rgb}{0.7,0.0,0.4}\definecolor{linkorange}{rgb}{0.7,0.1,0.0}\AtBeginDocument{\hypersetup{
linkcolor=linkblue,
citecolor=linkorange,
urlcolor=urlblue}}\definecolor{urlblue}{rgb}{0,0,0.9}\definecolor{linkblue}{rgb}{0,0,.8}\definecolor{linkgreen}{rgb}{0,0.45,0}\definecolor{linkpurple}{rgb}{0.7,0.0,0.4}\definecolor{linkorange}{rgb}{0.7,0.1,0.0}
\AtBeginDocument{\hypersetup{
linkcolor=linkblue,
citecolor=linkorange,
urlcolor=urlblue}}

\begin{document}
\newcommand{\newc}{\newcommand}

\newc{\ben}{\begin{eqnarray}}
\newc{\een}{\end{eqnarray}}
\newc{\be}{\begin{equation}}
\newc{\ee}{\end{equation}}
\newc{\ba}{\begin{eqnarray}}
\newc{\ea}{\end{eqnarray}}
\newc{\brk}[1]{\left(#1\right)}
\newc{\D}{\partial}
\newc{\rH}{{\rm H}}
\newc{\vp}{\varphi}
\newc{\rd}{{\rm d}}
\newc{\pa}{\partial}
\newc{\Mpl}{M_{\rm Pl}}
\newc{\ds}{\displaystyle}

\newcommand{\ma}[1]{\textcolor{magenta}{#1}}
\newcommand{\cy}[1]{\textcolor{cyan}{#1}}
\newcommand{\ST}[1]{\textcolor{blue}{[ST:~#1]}}
\newcommand{\re}[1]{\textcolor{red}{#1}}

\newcommand{\dd}{\textrm{d}}

\newcommand{\mq}[1]{\textcolor{blue}{#1}}
\newcommand{\rs}[1]{\textcolor{Green}{#1}}
\newcommand{\rsc}[1]{\textcolor{olive}{Comment by RS: #1}}

\begin{flushright}
\end{flushright}

\preprint{WUCG-23-04}
\preprint{ET-0137A-23}

\title{Constraining Horndeski theory with gravitational waves from coalescing binaries}

\author{Miguel Quartin$^{1,2,3,4}$, Shinji Tsujikawa$^{5}$, Luca Amendola$^{1}$, Riccardo Sturani$^{4,6}$}

\affiliation{
$^{1}$Institute of Theoretical Physics, Heidelberg University,
Philosophenweg 16, 69120 Heidelberg, Germany\\
$^{2}$Instituto de Fisica, Universidade Federal do Rio de Janeiro, 21941-972, Rio de Janeiro, RJ, Brazil\\
$^{3}$Observatório do Valongo, Universidade Federal do Rio de Janeiro, 20080-090, Rio de Janeiro, RJ, Brazil\\
$^{4}$PPGCosmo, Universidade Federal do Espírito Santo, 29075-910, Vitória, ES, Brazil\\
$^{5}$Department of Physics, Waseda University, 3-4-1 Okubo, Shinjuku, Tokyo 169-8555, Japan\\
$^{6}$Instituto de Física Teórica, UNESP-Universidade Estadual Paulista \& ICTP South American Institute for  Fundamental Research,  São Paulo 01140-070, SP, Brazil}

\begin{abstract}
    In the broad subclass of Horndeski theories with a luminal speed of gravitational waves, we derive gravitational waveforms emitted from a compact binary by considering the wave propagation on a spatially flat cosmological background. A scalar field nonminimally coupled to gravity gives rise to hairy neutron star (NS) solutions with a nonvanishing scalar charge, whereas black holes (BHs) do not have scalar hairs in such theories. A binary system containing at least one hairy neutron star modifies the gravitational waveforms in  comparison to those of the BH-BH binary. Using the tensor gravitational waveforms, we forecast the constraints on a parameter characterizing the difference of scalar charges of NS-BH or NS-NS binaries for Advanced LIGO and Einstein Telescope. We illustrate how these constraints depend on redshift and signal-to-noise ratio, and on different possible priors.
    We show that in any case it is possible to constrain the scalar charge precisely, so that some scalarized NS solutions known in the literature can be excluded.
\end{abstract}

\date{\today}


\maketitle

\section{Introduction}
\label{introsec}

The advent of gravitational wave (GW) astronomy, ushered in by the epochal detection of {GW150914}~\cite{LIGOScientific:2016aoc}, and by the first neutron star merger GW170817 \cite{LIGOScientific:2017vwq}, is being followed by a host of present, or second generation interferometers \cite{TheLIGOScientific:2014jea,TheVirgo:2014hva,KAGRA:2020tym} and future projects, consisting of third generation \cite{Punturo:2010zz,Maggiore:2019uih,Evans:2021gyd,Branchesi:2023mws} and space interferometers~\cite{Seto:2001qf,Kawamura:2011zz,LISA:2017pwj}, to detect more GW signals to high signal-to-noise ratio. GWs may be originated by merging processes of a variety of astrophysical objects, as well as from stars rapidly rotating or undergoing cataclysmic processes like supernova explosions, or even appear as a form of the stochastic background of either astrophysical or cosmological origin. The new ground opened by these developments is virtually unlimited: GWs can be employed to test the physics of black holes (BHs), neutron stars (NSs), progenitor channels~\cite{Kruckow:2018slo}, cosmic expansion~\cite{Schutz:1986gp}, large-scale structures~\cite{Alfradique:2022tox}, inflationary models~\cite{Guzzetti:2016mkm}, and, naturally,
the deviation from General Relativity (GR).

This paper is devoted to testing a class of scalar-tensor gravity \cite{Fujii:2003pa} through its prediction of inspiraling gravitational waveforms. Although gravity can be modified in several ways, one particularly simple model, rich of phenomenology and passing current constraints, is provided by Horndeski theories with second-order Euler
equations of motion \cite{Horndeski:1974wa}
(see also Refs.~\cite{Deffayet:2010qz,Deffayet:2011gz,Kobayashi:2011nu,Charmousis:2011bf,Kase:2018aps,Kobayashi:2019hrl}). Once the speed of gravity is enforced to be luminal to satisfy the simultaneous observation of gamma ray burst GRB 170817A~\cite{Goldstein:2017mmi} and merger signal of GW170817~\cite{LIGOScientific:2017zic}, the Horndeski Lagrangian reduces to the relatively simple form~\cite{Kobayashi:2011nu,DeFelice:2011bh,Baker:2017hug,Creminelli:2017sry,Sakstein:2017xjx,Ezquiaga:2017ekz}
\begin{equation}
    L=G_2(\phi,X)-G_3(\phi,X)\square \phi+G_4(\phi)R\,,
\end{equation}
where $G_2$ and $G_3$ depend on the scalar field $\phi$ and its kinetic term $X$ and $G_4$ is a function of $\phi$ alone with $R$ being the Ricci scalar. The same Lagrangian has been shown to produce many observable effects in cosmology (see e.g., \cite{Amendola:2019laa}). Recently, one of the present authors studied the effects of such a subclass of Horndeski theories  on the inspiraling gravitational waveform arising from the emission and propagation of GWs \cite{Higashino:2022izi}, extending earlier works in  Refs.~\cite{1977ApJ...214..826W,Damour:1996ke,Will:1994fb,Shibata:1994qd,Harada:1996wt,Brunetti:1998cc,Berti:2004bd,Scharre:2001hn,Alsing:2011er,Yunes:2011aa,Chatziioannou:2012rf,Barausse:2015wia,McManus:2016kxu,Zhang:2017srh,Liu:2018sia,Berti:2018cxi,Tahura:2018zuq,Niu:2019ywx,Liu:2020moh,Renevey:2020tvr,Renevey:2021tcz}.

In the subclass of Horndeski theories mentioned above, the nonminimal coupling $G_4(\phi)R$ gives rise to NS solutions endowed with scalar hairs, while there are no hairy BH solutions known in the literature for
regular coupling functions $G_2(\phi,X)$, $G_3(\phi,X)$, $G_4(\phi)$~\cite{Hawking:1972qk,Bekenstein:1995un,Sotiriou:2011dz,Graham:2014mda,Faraoni:2017ock,Minamitsuji:2022vbi}. A typical example is Brans-Dicke (BD) theory \cite{Brans:1961sx} with a scalar potential $V(\phi)$, which is given by the Lagrangian
\begin{equation}
\label{eq:bulk_J}
    L_{\rm BD} = \big(1-6Q^2\big)e^{-2Q\phi/\Mpl} X -V(\phi)
    + \frac{\Mpl^2}{2}e^{-2Q\phi/\Mpl}R \,,
\end{equation}
where $\Mpl$ is the reduced Planck mass and $Q$ is a coupling constant related to the BD parameter $\omega_{\rm BD}$ as $2Q^2=(3+2\omega_{\rm BD})^{-1}$~\cite{Khoury:2003rn,Tsujikawa:2008uc}. Since the scalar field is coupled to matter through the nonminimal coupling $(\Mpl^2/2)e^{-2Q\phi/\Mpl}R$, it mediates fifth forces in the vicinity of local sources like NS and Sun. For $V(\phi)=0$, Solar-System tests of gravity put the bound $\omega_{\rm BD}>4.0 \times 10^4$~\cite{Will:2014kxa}, which translates to the limit $|Q|<2.5 \times 10^{-3}$. In the presence of the potential $V(\phi)$, such a constraint on $|Q|$ can be loosened by a chameleon mechanism \cite{Khoury:2003aq,Khoury:2003rn,Sanctuary:2010dao}. This is also the case for the cubic Galileon coupling $G_3 \propto X$, which screens fifth forces under a Vainshtein mechanism~\cite{Vainshtein:1972sx,Burrage:2010rs,DeFelice:2011th,Kimura:2011dc,Koyama:2013paa}. In strong gravity backgrounds like the vicinity of NSs, the GW observations will allow us to put further constraints on model parameters of BD theories.

If the nonminimal coupling $G_4$ is given by an even power-law function of $\phi$, a phenomenon called spontaneous scalarization of relativistic stars can occur on a strong gravitational background.  For example, the function $G_4(\phi)=(\Mpl^2/2)e^{-\beta \phi^2/(2\Mpl^2)}$ with a negative coupling $\beta$, which was originally exploited by Damour and Esposito-Farese \cite{Damour:1993hw,Damour:1996ke} for spherically symmetric NSs, leads to tachyonic instability of the general relativistic branch ($\phi=0$) toward a nonvanishing scalar field branch. Spontaneous scalarization can occur for $\beta \le -4.35$~\cite{Harada:1998ge,Novak:1998rk,Sotani:2004rq,Silva:2014fca,Barausse:2012da}, whose limit weakly depends on the equations of state of NSs. This nonperturbative phenomenon in strong gravity regimes can be probed by the GW observations containing at least one NS in the compact binary system. The binary pulsar measurements of the energy loss through dipolar radiation put a stringent bound $\beta \ge -4.5$ \cite{Freire:2012mg,Shao:2017gwu,Anderson:2019eay}, so it remains to be seen whether future GW data can further reduce the allowed ranges of $\beta$. We note that, in the presence of a higher-order scalar kinetic term $\mu_2 X^2$ in $G_2$, it should be possible to evade tight constraints from binary pulsar measurements by suppressing a scalar charge  \cite{Higashino:2022izi,Shibata:2022gec}.

One can quantify the amount of the scalar charge in terms of a dimensionless parameter
\begin{equation}
    \hat{\alpha}_I=\Mpl\, \frac{\rd \ln \hat{m}_I(\phi)}{\rd \phi} \,,
\end{equation}
where $\hat{m}_I$ is a $\phi$-dependent mass of the compact object in the Einstein frame, obtained from the Jordan frame Lagrangian of equation~\eqref{eq:bulk_J} by canonically normalizing the kinetic terms of gravity and of the scalar field.  In BD theories, this parameter is crudely related to $Q$ as $\hat{\alpha}_I \simeq Q (1-3w_c)$, where $w_c$ is the equation of state parameter at the center of star \cite{Higashino:2022izi}.
In the original spontaneous scalarization scenario of Damour and Esposito-Farese, $|\hat{\alpha}_I|$ can reach the order of $0.1$ for typical NS
equations of state \cite{Damour:1993hw}. The presence of the $\mu_2 X^2$ term in $G_2$ can suppress the value of $|\hat{\alpha}_I|$ down to the order $0.01$ \cite{Higashino:2022izi}. For the gravitational waveform emitted from a compact binary during the inspiral stage, the difference of $\hat{\alpha}_I$ between two compact objects (labelled by $A$ and $B$), which we denote $\Delta \alpha=\hat{\alpha}_A-\hat{\alpha}_B$, appears in both phase and amplitude at leading-order in the post-Newtonian (PN) expansion~\cite{Alsing:2011er,Yunes:2011aa,Berti:2018cxi,Liu:2020moh,Higashino:2022izi}. In our subclass of Horndeski theories, $\Delta \alpha$ does not generally vanish for either NS-BH binaries (henceforth NSBH) or NS-NS binaries (henceforth BNS). Hence it is possible to probe the amount of scalar charges carried by NSs from the GW observations.

A natural scheme to confront gravitational theories
with observations is the \emph{parametrized post-Newtonian} (PPN) framework~\cite{Blanchet:2013haa,Will:2014kxa}. In this framework, the relativistic potential between two (or more) bodies are expanded in powers of the relative velocity $v$ with respect to the speed of light $c$ (we will use the unit $c=1$ throughout the paper), where $v$ is related to the
Newton's constant $G$, total binary mass $M$, and orbital radius $r$ via Kepler law $v^2\sim G M/r$. In the PPN framework at each perturbative order new phenomenological parameters appear, which on one hand can be computed theoretically in terms of the fundamental parameters of gravitational
theories and astrophysical properties of the source, and on the other can be directly checked with observations. Any modified gravitational theories should pass observational constraints arising from binary pulsar orbit
evolution \cite{Kramer:2021jcw}, gravitational waveforms~\cite{LIGOScientific:2018dkp,LIGOScientific:2021sio}, and Solar System tests of gravity  \cite{Will:2014kxa}.

For non- or moderately-relativistic sources it is convenient to expand the source terms in multipoles, and in GR, conservation laws allow radiation only at quadrupolar, or higher multipole level. This is not the case for scalar radiation, which can occur at \emph{dipolar}
level~\cite{Lang:2013fna,Mirshekari:2013vb,Lang:2014osa,Sennett:2016klh,Bernard:2018hta,Bernard:2022noq}. Hence it can appear at $v^{-2}$, or $-1$PN, order with respect to GR. Such term has been tested in a phenomenological manner (i.e., without a direct connection to a specific theory) by the LIGO/Virgo/KAGRA collaboration~\cite{LIGOScientific:2018dkp,LIGOScientific:2021sio}, which compares the observed waveform \emph{phase} evolution with the one predicted by modified gravity theories. However, for the wide class of theories gathered under the Horndeski umbrella, the extra scalar field affects both amplitude and phase, and, more importantly, the main additional scalar-tensor source parameter is correlated to other astrophysical parameters, making its constraint nontrivial.

In general, when comparing theoretical predictions with GW detections, a privileged role is played by the phase of the gravitational waveform.
This is because the output of GW observatories is processed via \emph{matched-filtering}~\cite{mf,LIGOScientific:2019hgc}, which is
particularly sensitive to the phase of the GW signal. We then investigate in the present work how the detected GW signal is affected by the GR deviation mostly represented by $\Delta \alpha$, and how well the scalar charge can be constrained by analyzing the signal produced by inspiraling binaries.

The scalar charge can be constrained through pulsar timing \cite{2020PhRvD.101j4011N}, which imposes a very strong bound. This method is however limited to Milky Way sources and to compact star binaries (in which case, as we show below, the scalar-tensor modification is expected to be suppressed), while here we include generic binary systems, with mixed NSBH having the highest signal. Moreover, since  pulsar timing tests nonrelativistic compact stars with $v\lesssim 10^{-3}$, there is little or no correlation with higher-order PN parameters, which is the opposite case of GW measurements, where sources with stellar masses or larger can be followed until merger when $v\lesssim 1$.

In this work, we forecast observational impacts of the inspiral gravitational waveform in the subclass of Horndeski theories on GW detectors of second generation (2G) like Advanced LIGO (aLIGO)~\cite{TheLIGOScientific:2014jea} and of third generation (3G), like the Einstein Telescope (ET)~\cite{Punturo:2010zz,Maggiore:2019uih,Branchesi:2023mws}. For this purpose, we provide for the first time the most general expression of tensor waveforms on a spatially flat cosmological background. In these theories the speed of gravity is luminal, but the nonminimal coupling gives rise to a friction term in the GW equation of motion. This leads to a  GW propagation different from that of light~\cite{Saltas:2014dha,Nishizawa:2017nef,Arai:2017hxj,Belgacem:2017ihm,Tsujikawa:2019pih}, whose property was used to forecast or constrain the running Planck mass on cosmological scales~\cite{Lombriser:2015sxa,Amendola:2017ovw,Zhao:2018gwk,Belgacem:2018lbp,Ezquiaga:2018btd,Lagos:2019kds,DAgostino:2019hvh}.

Performing a Fisher Matrix (FM) analysis, we  determine the expected uncertainty on GR and scalar-tensor parameters by the detection of either BNS and NSBH coalescences.
In particular, we show that it is possible to constrain the scalar charge from a coherent analysis of the GW phase from NSBH events to 0.0004 (0.0001) absolute level with 2G (3G) detectors. We investigate the relative importance of the Horndeski scalar charge modifications on different terms of the waveform and show that both the GW amplitude and the sub-dominant phase corrections are both quantitatively irrelevant to determine the uncertainty on the main GW parameters. We also explore how precision correlates with signal-to-noise level in the detection and with the choice of spin priors.

\section{Gravitational waveforms in scalar-tensor theories}
\label{Minsec}

We consider a subclass of Horndeski theories \cite{Horndeski:1974wa}
given by the action
\be
{\cal S}=
\int {\rm d}^4 x \sqrt{-g}
\left[ G_2(\phi,X)-G_3(\phi,X) \square \phi
+G_4(\phi)R
\right]+{\cal S}_m\,,
\label{action}
\ee
where $g$ is the determinant of metric tensor $g_{\mu\nu}$, $G_2$ and $G_3$ are functions of $\phi$ and $X\equiv -(1/2)\nabla^{\mu}\phi \nabla_{\mu}\phi$, with the covariant derivative operator denoted by $\nabla_{\mu}$, $\square \equiv g^{\mu \nu} \nabla_{\mu} \nabla_{\nu}$ is the d'Alembertian, and $G_4$ is given by
\be
G_4(\phi)=\frac{\Mpl^2}{2}F(\phi)\,,
\label{G4def}
\ee
where $\Mpl$ is the reduced Planck mass, and $F$ is a dimensionless
function of $\phi$.
We assume that the matter fields described by the action
${\cal S}_m$ are minimally coupled to gravity. In theories given by the action (\ref{action}) the speed of gravitational waves $c_t$ is equivalent to that of light \cite{Kobayashi:2011nu,DeFelice:2011bh}, so they automatically satisfy the observational bound on $c_t$ from the GW170817 event \cite{LIGOScientific:2017zic} together with the
electromagnetic (EM) counterpart~\cite{Goldstein:2017mmi}.

The nonminimal coupling $G_4(\phi) R$ can give rise to NS solutions endowed with scalar hair \cite{Damour:1993hw,Damour:1996ke}. Since $R$ is related to the matter energy-momentum tensor through gravitational equations of motion, matter fields
are affected by the scalar field through the coupling $F(\phi)R$. In this case, the Arnowitt-Deser-Misner (ADM) mass of NS depends on the scalar field $\phi$.
In theories given by the action (\ref{action}), it is known that there are no static and spherically
symmetric BH solutions with scalar hairs for regular
coupling functions $G_2(\phi,X)$, $G_3(\phi,X)$, and $G_4(\phi)$ \cite{Hawking:1972qk,Bekenstein:1995un, Sotiriou:2011dz,Graham:2014mda,Faraoni:2017ock,
Gomes:2020lvs,Minamitsuji:2022vbi}. We are mostly interested in a compact binary system containing at least one NS to probe signatures of the modification of gravity.
We use the labels $I=A,B$ for the field-dependent ADM masses $m_I(\phi)$ to deal with the binary system as a collection of two point-like particles, modeling the matter sources as massive objects with a fixed world-line. Then, the matter action is given by~\cite{Eardley1975}
\be
\label{eq:wl}
{\cal S}_m=
-\sum_{I=A,B} \int m_I(\phi){\rm d}\tau_I\,,
\ee
where $\tau_I$ is the proper time along
the world line of particle $I$.

A conformal transformation of metric tensor such that
$\hat{g}_{\mu \nu}=F(\phi) g_{\mu \nu}$ has
the virtue of diagonalizing the kinetic terms of the scalar and gravitational fields \cite{Fujii:2003pa}.
In the transformed Einstein frame,
the ADM mass of particle $I$ is given by
$\hat{m}_I (\phi)=m_I(\phi)/\sqrt{F(\phi)}$ \cite{Damour:1992we}.
On the asymptotically-flat background
we introduce the following
quantity directly related to a scalar charge \cite{Higashino:2022izi}
\be
\hat{\alpha}_I \equiv \frac{\Mpl \hat{m}_{I,\phi}}{\hat{m}_I}
\bigg|_{\phi=\phi_0}
=\frac{\Mpl m_{I,\phi}}{m_I}\bigg|_{\phi=\phi_0}
-\frac{1}{2}g_4\,,
\label{eq:alpha}
\ee
where $\phi_0$ is an asymptotic value of $\phi$,
$g_4 \equiv \Mpl F_{,\phi}/F|_{\phi=\phi_0}$, and
we used the notation $\hat{m}_{I,\phi}={\rm d}\hat{m}_I/{\rm d}\phi$.
For a scalarized NS we have $\hat{\alpha}_I \neq 0$, while $\hat{\alpha}_I=0$ for no-hair BHs.

\subsection{Minkowski background}

In theories given by the action (\ref{action}), the propagation of
GWs from the binary to an observer was addressed in
Ref.~\cite{Higashino:2022izi} on the Minkowski background given the line element $\rd s^2=-\rd t^2+\delta_{ij}\rd x^i \rd x^j$.
On this background,
the metric tensor $g_{\mu \nu}$ and the scalar field $\phi$ can be expanded as
$g_{\mu \nu}=\eta_{\mu \nu}+h_{\mu \nu}$ and
$\phi=\phi_0+\varphi$, where $\eta_{\mu \nu}={\rm diag}\,(-1,1,1,1)$
and $\phi_0$ are the background values and
$h_{\mu \nu}$ and $\varphi$ are the perturbed
quantities. Time-domain solutions to $h_{\mu \nu}$
and $\varphi$ were derived in Ref.~\cite{Higashino:2022izi}
under the PN expansion up to
quadrupole order (see also Refs.~\cite{Alsing:2011er,Berti:2012bp,Sagunski:2017nzb,
Zhang:2017srh,Liu:2018sia,Niu:2019ywx,Liu:2020moh}
for related works).
For the validity of the PN expansion,
we require that scalar field nonlinear derivative interactions arising from the cubic Lagrangian
$G_3(X) \square \phi$ are suppressed outside the
source. This means that, for the Vainshtein
radius $r_V$ much larger than the star radius $r_s$,
the gravitational waveform derived in Ref.~\cite{Higashino:2022izi} loses its validity.
When the Lagrangian $G_3(X) \square \phi$ is present,
we assume that $r_V$ is less than $r_s$.
In this case, the screening of fifth forces can occur
only inside the star.

Besides the traceless-transverse polarizations
$h_{+}$ and $h_{\times}$ of GWs, the scalar field
perturbation $\varphi$ gives rise to breathing and longitudinal polarizations, which are denoted as $h_{b}$ and $h_{L}$ respectively. Note that $h_b$ is isotropic in the plane transverse to the propagation direction
defining the longitudinal direction.
Provided that $|\hat{\alpha}_I| \ll 1$, the amplitudes
of $h_{b}$ and $h_{L}$ are suppressed relative to those of
$h_{+}$ and $h_{\times}$.
In this paper, we will focus on the gravitational waveforms
of $h_{+}$ and $h_{\times}$ alone.
We caution, however, that the breathing and longitudinal modes
need to be taken into account for complete constraints on scalar-tensor
theories \cite{Takeda:2018uai,Moretti:2019yhs,Takeda:2020tjj,Takeda:2021hgo}.

Since we are interested in a light scalar field relevant to
the late-time cosmology, we assume that
the scalar field mass $m_s$ is much smaller than
the typical orbital frequency
$\omega \approx 10^{-13}$~eV$/\hbar$
of the binary system.
In such cases, the scalar radiaton during
the inspiral stage
leads to modified gravitational waveforms of
$h_{+}$ and $h_{\times}$ in comparison to those
in GR \cite{Higashino:2022izi}.

We define the following perturbed quantity
\be
\label{def:theta}
\theta_{\mu \nu} \equiv h_{\mu \nu}
-\frac{1}{2} \eta_{\mu \nu} h
-\eta_{\mu \nu} g_4 \frac{\varphi}{\Mpl}\,,
\ee
together with the trace
$h \equiv \eta^{\mu \nu} h_{\mu \nu}$.
Choosing the transverse gauge condition
$\nabla^{\nu} \theta_{\mu \nu}=0$,
the linearized equation for $\theta_{\mu \nu}$
on the Minkowski background
is given by
\be
\left( \frac{\partial^2}{\partial t^2}-\nabla^2 \right)
\theta_{\mu \nu}=\frac{2T_{\mu \nu}^{(1)}}{\Mpl^2 F(\phi_0)}\,,
\label{thetaeq}
\ee
where
$T_{\mu \nu}^{(1)}$ is the perturbed
energy-momentum tensor at first order.
The solution to Eq.~(\ref{thetaeq}) at spacetime point
$x^{\alpha}=(t,{\bm x})$ can be expressed as
\be
\theta_{\mu \nu} (x^{\alpha})
=\frac{1}{2\pi \Mpl^2 F(\phi_0)} \int \rd^3 x'
\frac{T_{\mu \nu}^{(1)}(t-|{\bm x}-{\bm x}'|, {\bm x}')}
{|{\bm x}-{\bm x}'|}\,.
\ee
At a space position ${\bm x}$ with the comoving distance $d_s$ far away from the center of the binary, we use the approximation $|{\bm x}-{\bm x}'|=d_s-{\bm x}' \cdot {\bm n}$ and expand $T_{\mu \nu}^{(1)}(t-|{\bm x}-{\bm x}'|, {\bm x}')$
about the retarded time $t-d_s$.
Then, the leading-order solution to the spatial components of $\theta^{\mu \nu} (x)$ is given by the quadrupole formula
\be
\theta^{ij} (x^{\alpha})=\frac{1}{4\pi \Mpl^2 F(\phi_0) d_s}
\frac{\pa^2}{\pa t^2} \sum_{I=A,B} m_I
x_I^i x_I^j\,,
\label{thetaij}
\ee
where $x_I^i$ ($I=A,B$) are the positions of
point sources $A$ and $B$.
The effective gravitational coupling appearing in
the Newtonian equations of motion for $A$ and $B$
is given by \cite{Higashino:2022izi}
\be
G=G_{*0} \left( 1+\delta_0 \right)\,,
\ee
where
\be
G_{*0} \equiv \frac{1}{8 \pi \Mpl^2 F(\phi_0)}\,,\qquad
\delta_0 \equiv 4 \kappa_0 \hat{\alpha}_A
\hat{\alpha}_B\,,\qquad
\kappa_0 \equiv
\frac{F(\phi_0)}{2\zeta_0}\,,\qquad
\zeta_0 \equiv G_{2,X}-2G_{3,\phi}
+\frac{3\Mpl^2 F_{,\phi}^2}{2F} \biggr|_{\phi=\phi_0}\,.
\ee
If the two compact bodies have nonvanishing
scalar charges $\hat{\alpha}_A$ and $\hat{\alpha}_B$,
the Newtonian gravitational coupling $G$ is affected by
them through the quantity $\delta_0$.
For the binary system containing a no hair BH,
we have $\delta_0=0$ and hence $G$ is equivalent
to $G_{*0}$.

For a quasicircular orbit of the binary system with
the relative distance $r$, the quadrupole
formula (\ref{thetaij}) further reduces to
\be
\theta^{ij}(x)=\frac{G \mu m}{2\pi \Mpl^2
F(\phi_0) r d_s} \left( \hat{v}^i \hat{v}^j
-\hat{r}^i \hat{r}^j \right)\,,
\label{thetaijdef}
\ee
where $\hat{v}^i$ and $\hat{r}^i$ are unit vectors
in the directions of velocity and positions,
respectively, and
\be
\mu \equiv \frac{m_A m_B}{m_A+m_B}\,,\qquad
m \equiv m_A+m_B\,.
\ee
with the relative circular velocity $v=(G m/r)^{1/2}$.

Let us consider a Cartesian coordinate system $(x,y,z)$, whose origin O coincides with the center of mass of the binary system which is supposed to lie in the $(x,y)$ plane. We discuss the propagation of GWs from O to the observer (point P) in the $(y,z)$ plane with an angle
$\iota$ inclined from the $z$ axis. The unit vector in the direction of $\overrightarrow{\rm OP}$ is given by $n^i=(0, \sin \iota, \cos \iota)$.
The quasicircular motion of the binary system is confined on the $(x,y)$ plane with an angle $\Phi$ inclined from the $x$ axis, so that
$\hat{r}^i=(\cos \Phi, \sin \Phi, 0)$ and $\hat{v}^i=(-\sin \Phi, \cos \Phi, 0)$. In this configuration, the traceless and transverse (TT)
components of $\theta_{ij}$ (denoted as $\theta_{ij}^{\rm TT}$)
propagating along the direction of $n^i$ have two polarized modes
$h_{+}=\theta_{11}^{\rm TT}=-\theta_{22}^{\rm TT}$ and $h_{\times}=\theta_{12}^{\rm TT}=\theta_{21}^{\rm TT}$. Under the quadrupole approximation, they are given, respectively, by
\ba
h_{+}(t) &=&
-(1+\delta_0)^{2/3}
\frac{4(G_{*0} {\cal M})^{5/3} \omega^{2/3}}{d_s}
\frac{1+\cos^2 \iota}{2} \cos(2\Phi)\,,\label{hp}\\
h_{\times}(t)
&=&-(1+\delta_0)^{2/3}
\frac{4(G_{*0} {\cal M})^{5/3} \omega^{2/3}}{d_s}
\cos \iota \sin(2\Phi)\,,\label{ht}
\ea
where ${\cal M} = \mu^{3/5} m^{2/5} = m \eta^{3/5}$ with $\eta \equiv \mu/m$ being the symmetric mass ratio, $\omega=v/r$ is the orbital frequency, and the orbital phase $\Phi=\omega \times (t-d_s)$ for perfectly circular motion.  Equations (\ref{hp}) and (\ref{ht}) correspond to gravitational  waveforms in the time domain at a distance $d_s$ from  the center of a binary.

The output of a detector $a$ (denoted as $h_a$ which is given by a single time series) is obtained by combining linearly the two polarizations $h_{a,\genfrac{}{}{0pt}{3}{+}{\times}}$ via  the \emph{pattern functions} $F_{\genfrac{}{}{0pt}{3}{+}{\times}}$
\be
\label{eq:hdet}
h_a=F_{a+}h_++F_{a\times} h_\times\,,
\ee
where $F_{a,\genfrac{}{}{0pt}{3}{+}{\times}}$ depend on sky location and on the polarization angle $\psi_a$ relative to detector $a$. Here, $\psi_a$ can be considered as the third Euler angle, besides $\iota$ and $\Phi$, relating the frame of the source to the one in which the radiation is propagating along the $\hat z$ axis \cite{Apostolatos:1994mx,deSouza:2023gjv}, and the $(\hat{x},\hat{z})$ plane contains the unit vector perpendicular to the detector $a$.
The angle $\psi_a$ parameterizes  a rotation in the plane perpendicular to the propagation direction of GWs. For a source sky location identified by polar angles $\delta_a,\gamma_a$, one has the pattern functions
$F_{a,\genfrac{}{}{0pt}{3}{+}{\times}} = \cos(2\psi) f_{a,\genfrac{}{}{0pt}{3}{+}{\times}}\mp \sin(2\psi) f_{a,\genfrac{}{}{0pt}{3}{\times}{+}}$ where for convenience we defined 
\ba
\ds f_{a+} &=&
\ds \sin(\Omega_a)
\frac{1+\cos^2(\delta_a)}{2}\cos(2\gamma_a)\,,\\
\ds f_{a\times} &=&
\ds \sin(\Omega_a)\cos(\delta_a)\sin(2\gamma_a)\,,
\ea
where we have allowed the $a$ interferometer arms to have a generic angular opening $\Omega_a$. One has $\Omega=\pi/2$ for 2G detectors, but there is the possibility that 3G ones will involve a triangular interferometer consisting of three observatories each with $\Omega=\pi/3$~\cite{Punturo:2010zz,Branchesi:2023mws}.

Taking into account the energy losses trough gravitational and scalar radiation, the orbital frequency $\omega$ becomes nonlinear in time, and its expression in frequency domain under a stationary phase approximation will be derived in Sec.~\ref{cosmosec}. While the above discussion is based on the propagation of GWs on the Minkowski background, we will extend the analysis to the derivation of frequency-domain gravitational waveforms on the cosmological background.

\subsection{Cosmological background}
\label{cosmosec}
The line element of a spatially-flat cosmological background is given by
\be
\rd s^2=-\rd t^2+a^2(t) \delta_{ij} \rd x^i \rd x^j\,,
\label{metric}
\ee
where $a(t)$ is a time-dependent scale factor. The redshift $z$ of the binary source can be expressed as  $1+z=a(t_0)/a(t_s)$, {where $t_0$ and $t_s$ are respectively the times of GW emission and detection.}

The frequency of GWs measured by the observer,  $f_0$, is related to the corresponding  frequency measured in the source frame, $f_s$, as
\be
f_0=(1+z)^{-1} f_s\,.
\label{fre}
\ee
The luminosity distance travelled by light from redshift $z>0$ to the observer ($z=0$) is expressed as
\be
d_L(z)=(1+z) a(t_0) d\,,
\ee
where $d$ is a comoving distance defined by
\be
d=\frac{1}{a(t_0)}
\int_0^z \frac{\rd \tilde{z}}{H(\tilde{z})}\,.
\ee
Here, $H=\dot{a}/a$ is the Hubble expansion rate, with a dot being the derivative with respect to $t$.

In the local wave zone where the distance from the source is sufficiently large to have the $1/d_s$ behaviour in $h_+$ and $h_{\times}$, but still the effect of cosmic expansion on the propagation of GWs is negligible, we can replace $d_s$ in Eqs.~(\ref{hp}) and (\ref{ht}) with $a(t_s)d$.
In this regime, the time-domain solutions
of $h_{+}$ and $h_{\times}$ at time $t=t_s$
are given by
\ba
    h_{+}(t_s) &=&
    -h_c(t_s) \frac{1+\cos^2 \iota}{2} \cos(2\Phi)\,,\label{hplus}\\
    h_{\times}(t_s)
    &=&
    -h_c(t_s) \cos \iota \sin(2\Phi)\,,\label{hcross}
\ea
where $\Phi=\omega_s \times (t_s-d_s)$ with $\omega_s=\omega(t_s)$, and
\be
    \label{eq:ha}
    h_c(t_s)=(1+\delta_s)^{2/3} \frac{4(G_* {\cal M})^{5/3}
    \omega_s^{2/3}}{a(t_s)d}\,,
\ee
with
\be
    G_* \equiv \frac{1}{8 \pi \Mpl^2 F(\phi_s)}\,,
    \qquad
    \delta_s \equiv 4 \kappa_s \hat{\alpha}_A
    \hat{\alpha}_B\,,\qquad
    \kappa_s \equiv
    \frac{F(\phi_s)}{2\zeta_s}\,,\qquad
    \zeta_s \equiv G_{2,X}-2G_{3,\phi}
    +\frac{3\Mpl^2 F_{,\phi}^2}{2F} \biggr|_{\phi=\phi_s}\,.
    \label{Gstar}
\ee
Note that $\phi_s$ is the field value at the source.

We denote the time-dependent background scalar field as $\phi=\phi(t)$. In Fourier space with the comoving wavenumber $k$, the equations of motion for the TT modes $h_{\lambda}$ (with $\lambda=+, \times$) away from the source are~\cite{Kobayashi:2011nu,DeFelice:2011bh,Saltas:2014dha,Nishizawa:2017nef,
Kase:2018aps,Tsujikawa:2019pih}
\be
\label{heqcos}
\ddot{h}_{\lambda}+\left( 3+\alpha_{\rm M} \right) H
\dot{h}_{\lambda}+\frac{k^2}{a^2}h_{\lambda}=0\,,
\ee
where
\be
\alpha_{\rm M}=\frac{\dot{F}}{HF}
=\frac{F_{,\phi}}{HF} \dot{\phi}(t)\,.
\label{aM}
\ee

So long as the scalar field evolves over
the cosmological time, $\alpha_{\rm M}$ is nonvanishing.
We note that there are some observational bounds on $\alpha_{\rm M}$ already set
in a phenomenological manner \cite{Arai:2017hxj,Battye:2018ssx,LISACosmologyWorkingGroup:2019mwx,Mastrogiovanni:2020gua,DAgostino:2019hvh}. We define the rescaled GW field
\be
    \hat{h}_{\lambda} \equiv a_{\rm GW}h_{\lambda}\,,
    \qquad {\rm with} \qquad
    a_{\rm GW}=a \sqrt{\frac{F(\phi)}{F(\phi_0)}}\,,
\ee
where $\phi_0$ is today's background field value.
Then, Eq.~(\ref{heqcos}) can be expressed in the form
\be
    \hat{h}_{\lambda}''+\left( k^2 -\frac{a_{\rm GW}''}
    {a_{\rm GW}} \right)\hat{h}_{\lambda}=0\,,
    \label{thl}
\ee
where a prime represents the derivative with
respect to $\eta=\int a^{-1}\rd t$.
For the perturbations deep inside the Hubble radius,
the term $a_{\rm GW}''/a_{\rm GW}$
can be ignored relative to $k^2$.
Then, the solution to Eq.~(\ref{thl}) is given by
$\hat{h}_{\lambda}=A e^{\pm i k \eta}$, where
$A$ is a constant.
Since the amplitude of $h_{\lambda}=\hat{h}_{\lambda}/a_{\rm GW}$
is proportional to $1/a_{\rm GW}$,
today's value of $h_c(t_0)$ is given by
\be
h_c(t_0)=\frac{a_{\rm GW}(t_s)}{a_{\rm GW}(t_0)}
h_c(t_s)=\frac{a_{\rm GW}(t_s)}{a(t_s)}
(1+\delta_s)^{2/3} \frac{4(G_* {\cal M})^{5/3}
\omega_s^{2/3}}{a(t_0)d}\,,
\label{hct}
\ee
where we used $a_{\rm GW}(t_0)=a(t_0)$ in the second equality.
We define the GW distance \cite{Nishizawa:2017nef,Belgacem:2017ihm,Amendola:2017ovw,DAgostino:2019hvh,Matos:2022uew}
\be
d_{\rm GW}(z)=d_L(z) \frac{a(t_s)}{a_{\rm GW}(t_s)}
=d_L(z) \sqrt{\frac{F(\phi_0)}{F(\phi_s)}}\,,
\label{dua}
\ee
where $\phi_s$ is the background field value at the source.
Then, one can express Eq.~(\ref{hct}) in the form
\be
h_c(t_0)=(1+z)(1+\delta_s)^{2/3} \frac{4(G_* {\cal M})^{5/3}
\omega_s^{2/3}}{d_{\rm GW}(z)}
=(1+\delta_s)^{2/3} \frac{4\big(G_* \widetilde{\cal M}\big)^{5/3}
\omega_0^{2/3}}{d_{\rm GW}(z)}\,,
\ee
where
\be
\widetilde{\cal M}=(1+z){\cal M}=(1+z)\mu^{3/5} m^{2/5}\,,\qquad
\omega_0=(1+z)^{-1} \omega_s\,.
\ee
Here $\widetilde{\cal M}$ is the redshifted chirp mass, and $\omega_0$ is the orbital frequency measured by the observer's clock. The time-domain TT polarized modes measured by the observer are
$h_{+}(t_0)=-h_c(t_0) [(1+\cos^2 \iota)/2] \cos(2\Phi)$ and $h_{\times}(t_0)=-h_c(t_0) \cos \iota \sin(2\Phi)$.

The binary system loses its energy through the
gravitational and scalar radiations.
This leads to the increase of the orbital
frequency $\omega$.
The time-domain solutions of $h_{\lambda}(t)$
can be mapped to the frequency-domain solutions
$\tilde{h}_{\lambda} (f)$ under
the Fourier transformation
\be
\label{eq:spa}
\tilde{h}_{\lambda} (f_0)=
\int {\rm d}t_0\,h_{\lambda}(t_0)\,
e^{i \cdot 2 \pi f_0 t_0}=
(1+z) \int \rd t_s\,h_{\lambda}(t_0)
e^{i \cdot 2 \pi f_s t_s}\,,
\ee
where $f_{0,s}$ denote the GW quadrupolar frequency at detection and source, respectively.
In the exponent of Eq.~(\ref{eq:spa}),
we used the approximation
$t_0=(1+z)t_s$, which is only approximately true as
$\rd t_0=(1+z) \rd t_s$ and Eq.~(\ref{fre})
exactly holds.
This can induce a relative error of order ${\cal O}(H_0\Delta t)$ \cite{Seto:2001qf,Nishizawa:2011eq,Bonvin:2016qxr},
being $\Delta t$ the time duration of the GW signal,
but we will ignore such a small error here.

For the $\lambda=+$ mode, we have
\be
\tilde{h}_{+} (f_0)=
-(1+z)^{2}(1+\delta_s)^{2/3}
\frac{(G_* {\cal M})^{5/3}}{d_{\rm GW}(z)}
(1+\cos^2 \iota)\,
e^{i \cdot 2 \pi f_s d_s}
\int {\rm d}t_s\, \omega_s^{2/3}
\left[ e^{i(2\Phi(t_s)+2 \pi f_s t_s)}
+e^{i(-2\Phi(t_s)+2 \pi f_s t_s)} \right]\,.
\label{hp2}
\ee
The second term in the square bracket of
Eq.~(\ref{hp2}) has
a stationary phase point at $\omega(t_{s*})=\pi f_s$.
Expanding $\Phi(t_s)$ around $t_{s}=t_{s*}$ as
$\Phi(t_s)=\Phi(t_{s*})+\pi f_s (t_s-t_{s*})
+\dot{\omega}(t_{s*})(t_s-t_{s*})^2/2$
and dropping the fast oscillating mode
in Eq.~(\ref{hp2}), it follows that
\be
\tilde{h}_{+} (f_0)=
-(1+z)^{2}(1+\delta_s)^{2/3} \frac{(G_* {\cal M})^{5/3}}
{d_{\rm GW}(z)}
(1+\cos^2 \iota)\,\omega_s(t_{s*})^{2/3}
\sqrt{\frac{\pi}{\dot{\omega_s}(t_{s*})}}
e^{i \Psi_+}\,,
\label{hp4}
\ee
where $\Psi_{+}=2\pi f_s (t_{s*}+d_s) -2\Phi(t_{s*})-\pi/4$. Note that the distance $d_s$ appears in $\Psi_{+}$ because we have integrated the phase term $\Phi=\omega_s \times (t_s-d_s)$ with respect to $t_s$. This constant distance $d_s$ in $\Psi_{+}$ can be absorbed into $t_{s*}$ by shifting the origin of time, such that $t_{s*}+d_s \to t_{s*}$. Then, it follows that
\be
\Psi_{+}=2\pi f_s t_{s*}
-2\Phi(t_{s*})-\frac{\pi}{4}\,.
\label{Psip}
\ee
Analogous to the discussion above, the Fourier-transformed mode of  $h_{\times}(f_0)$ yields
\be
    \tilde{h}_{\times} (f_0)=
    -2(1+z)^2(1+\delta_s)^{2/3}
    \frac{(G_* {\cal M})^{5/3}}
    {d_{\rm GW}(z)}
    (\cos \iota)\,\omega_s(t_{s*})^{2/3}
    \sqrt{\frac{\pi}{\dot{\omega}_s(t_{s*})}}
    e^{i \Psi_\times}\,,
    \label{hm4}
\ee
where
\be
\Psi_{\times}=\Psi_{+}+\frac{\pi}{2}\,.
\ee

Taking into account the leading quadrupole formula emission for the gravitational spin-2 field and the leading dipolar emission for the scalar field, the time derivative of $\omega_s$ is given by \cite{Higashino:2022izi}
\be
\dot{\omega}_s=\frac{96}{5} (G_* {\cal M})^{5/3} \omega_s^{11/3}
\left[ 1+\frac{5\kappa_{s} (\hat{\alpha}_A-\hat{\alpha}_B)^2}
{24 (G_* m \omega_s)^{2/3}} \right]\,,
\label{omes}
\ee
where we ignored the term $\delta_s \approx \kappa_s \hat{\alpha}_I^2$
relative to the term $5\kappa_s (\hat{\alpha}_A-\hat{\alpha}_B)^2
/[24 (G_* m \omega)^{2/3}] \approx \kappa_s \hat{\alpha}_I^2/v^2$.
We define the critical time $t_{s\infty}$ at which $\omega$
grows sufficiently large, i.e., $\omega(t_{s\infty}) \to \infty$
with $\Phi_c=\Phi(t_{s\infty})$.\footnote{Solving Eq.~(\ref{omes}) for $\omega_s$, we find that $\omega_s$ diverges in a finite time $t_{s\infty}$, while $\Phi$ is finite at $t_{s\infty}$.}
The phase (\ref{Psip}) can be expressed as
\be
\Psi_{+}=2\pi f_s t_{s\infty}
-2\Phi_c-\frac{\pi}{4}
+\int_{\infty}^{\pi f_s} {\rm d} \omega_s \frac{2\pi f_s-2\omega_s}
{\dot{\omega}_s}\,.
\label{Psipl}
\ee

Substituting Eq.~(\ref{omes}) into Eqs.~(\ref{hp4}), (\ref{hm4}), and (\ref{Psipl}), we obtain the gravitational waveforms in the
frequency domain as
\ba
\tilde{h}_{+}(f_0) &=& -\tilde{h}_c (f_0)\frac{1+\cos^2 \iota}{2}
e^{i \Psi_{+}}\,,\\
\tilde{h}_{\times}(f_0) &=& -\tilde{h}_c (f_0) (\cos \iota)
e^{i \Psi_{\times}}\,,
\ea
where
\ba
    \tilde{h}_c (f_0)
    &=&\sqrt{\frac{5\pi}{24}} \frac{\big(G_* \widetilde{\cal M}\big)^{5/6}}
    {d_{\rm GW}(z)} \left( \pi f_0 \right)^{-7/6}
    \left[ 1-\frac{5 \kappa_s (\Delta \alpha)^2}{
    48(G_* \widetilde{m} \pi f_0)^{2/3}}
    \right]\,,\label{gwam}\\
    \Psi_{+}
    &=&
    \Psi_{\times}-\frac{\pi}{2}
    = 2\pi f_0 t_{c}-2\Phi_c-\frac{\pi}{4}
    +\frac{3}{128} \big(G_* \widetilde{\cal M} \pi f_0\big)^{-5/3}
    \left[ 1-\frac{5 \kappa_s (\Delta \alpha)^2}
    {42 (G_* \widetilde{m} \pi f_0)^{2/3}} \right]\,,
    \label{Psipl2}
\ea
with
\begin{equation}
    \Delta \alpha = \hat{\alpha}_A-\hat{\alpha}_B\,,
    \qquad \widetilde{m}=(1+z)m\,,\qquad
    t_{c}=(1+z) t_{s\infty}\,.
\end{equation}
The GW distance $d_{\rm GW}(z)$ appears in the denominator of the amplitude (\ref{gwam}). Note that $d_{\rm GW}(z)$ is different from $d_L(z)$ by the factor $\sqrt{F(\phi_0)/F(\phi_s)}$.

For scalarized NSs in the original Damour and Esposito-Farese scenario, the scalar charge can be as large as $|\hat{\alpha}_A|={\cal O}(0.1)$~\cite{Damour:1993hw,Niu:2021nic,Higashino:2022izi} (see also Appendix~\ref{app:Delta-alpha}). In such cases, the scalar charge gives a large contribution to $\tilde{h}_c (f_0)$ through the second term in the square bracket of Eq.~(\ref{gwam}). This correction can be more important than the one due to the change from $d_L$ to $d_{\rm GW}$. Moreover, the nonvanishing scalar charge modifies the phases $\Psi_+$ and $\Psi_{\times}$ in comparison to those in GR, so $\hat{\alpha}_A$ can be tightly constrained from the GW observations.

In the subclass of Horndeski theories given by the action (\ref{action}), the NS can have hairy solutions, while this is not the case for BHs. Then, for the NSBH, we can consider the case $\hat{\alpha}_A \neq 0$ and $\hat{\alpha}_B=0$, with $\hat{\alpha}_A$ quantifying the scalar charge carried by the NS. In BD theories with the nonminimal coupling  $F(\phi)=e^{-2Q \phi/\Mpl}$, extrapolating internal and external solutions of the scalar field up to the star radius gives the following approximate formula~\cite{Higashino:2022izi}
\be
\hat{\alpha}_A \simeq Q(1-3w_c)\,,
\label{alQre}
\ee
where $w_c$ is the equation of state parameter
at the center of NS.
Thus, $\hat{\alpha}_A$ depends on the nonminimal coupling as well as on the equation of state
inside the NS.

The gravitational waveforms derived above
belong to those in
the parameterized post-Einsteinian (ppE) framework
given by the general parametrization \cite{Yunes:2009ke,Cornish:2011ys,Chatziioannou:2012rf,Tahura:2018zuq}
\be
    \tilde{h}_{\lambda}(f_0)=
    \tilde{h}_{\lambda}^{\rm GR}(f_0)
    \left[ 1+\sum_{j=1} A_j
    \left( G_* \widetilde{\cal M} \pi f_0 \right)^{a_j/3} \right]
    \,\exp \left[ i \sum_{j=1}B_j
    \left( G_* \widetilde{\cal M} \pi f_0 \right)^{b_j/3} \right]\,,
    \label{PPE}
\ee
where $\tilde{h}_{\lambda}^{\rm GR}(f_0)$ are the $\lambda=+, \times$ polarizations in GR, and  $A_j$, $a_j$, $B_j$, $b_j$ are constants characterizing the deviation from GR. The leading-order correction to the amplitude (\ref{gwam})
arising from the modification of gravity  corresponds to the following ppE parameters
\be
a_1=-2\,,\qquad
A_1=-\frac{5}{48}\kappa_s
\left( \Delta \alpha \right)^2 \eta^{2/5}\,,
\ee
where $\eta \equiv \mu/m=(\widetilde{\cal M}/\widetilde{m})^{5/3}$.
The leading-order correction to the phase (\ref{gwam})
corresponds to
\be
    b_1=-7\,,\qquad
    B_1=-\frac{5}{1792} \kappa_s
    \left( \Delta \alpha \right)^2 \eta^{2/5}\,,
\ee
which can be regarded as the PN  correction at $-1$ order.
From the observed gravitational waveforms, we can put constraints on the ppE parameters $A_1$  and $B_1$. Since the GW observations are usually more sensitive to the phase rather than the amplitude, the parameter $B_1$ is expected to be more tightly constrained than $A_1$.
In particular, it is possible to place constraints on
the combination $\kappa_s (\Delta \alpha)^2$.

Let us consider theories given by
the coupling functions
\be
G_2=\left( 1-\frac{3\Mpl^2 F_{,\phi}^2}{2F^2}
\right) F(\phi)X\,,\qquad
G_3=0\,,
\label{G234}
\ee
with $G_4$ given by Eq.~(\ref{G4def}). BD theories \cite{Brans:1961sx} correspond to the nonminimal coupling $F(\phi)=e^{-2Q\phi/\Mpl}$, in
which case $G_2=(1-6Q^2)e^{-2Q\phi/\Mpl}X$.  GR with a canonical scalar field can be recovered by taking the limit $Q \to 0$. Theories of spontaneous scalarization advocated  by Damour and Esposito-Farese~\cite{Damour:1993hw,Damour:1996ke} correspond to the choice $F(\phi)=e^{-\beta \phi^2/(2\Mpl^2)}$. For the coupling functions (\ref{G234}), we have $\kappa_s=1/2$, in which case the observational bound on $B_1$  directly translates to $(\Delta \alpha)^2$. Taking into account the higher-order kinetic term $\mu_2 X^2$ in $G_2$ can suppress the scalar charge through kinetic screening inside the NS \cite{Higashino:2022izi,Shibata:2022gec}.
In this case, the field kinetic term appears in $\kappa_s$ as $\kappa_s=(1/2)(1+2\mu_2 X/F)^{-1}$.

We note that the scalar radiation besides the
gravitational radiation emitted from
a binary pulsar modifies the orbital period
in comparison to that in GR \cite{Damour:1991rd,Freire:2012mg,Shao:2017gwu}.
From the binary pulsar observations, there have been
constraints on the ppE parameters present in the
literature \cite{Yunes:2010qb,2020PhRvD.101j4011N,Kramer:2021jcw}.
For the $-1$PN corrections discussed above,
the measurements of PSR J0737-3039 give
the following bounds \cite{2020PhRvD.101j4011N}
\be
|A_1|< 6 \times 10^{-10}\,,\qquad
|B_1|< 2 \times 10^{-11}\,.
\label{albe1}
\ee
PSR J0737-3039 is composed of a binary with
the masses $m_A \simeq 1.338M_{\odot}$ and
$m_B \simeq 1.249M_{\odot}$ (where
$M_{\odot}$ is a solar mass),
in which case $\eta \simeq 0.245$.
{}From the two constraints (\ref{albe1}),
we obtain the following upper bound
\be
\sqrt{\kappa_s}\,|\Delta \alpha| \lesssim
10^{-4}\,.
\label{Dalbo}
\ee
In theories given by the coupling functions
(\ref{G234}), the bound (\ref{Dalbo})
translates to
$|\Delta \alpha| \lesssim 10^{-4}$.

Due to the large number of observed cycles, pulsar constraints are currently much stronger than those from GWs (see Sec.~\ref{foresec}), but they are  limited to very few sources and to the Milky Way. Many more sources at deep distances are expected in the next LIGO-Virgo-KAGRA runs, so that their combination will increase precision. Also, as discussed, NSBH mergers are expected to produce larger deviations from GR in this class of theories with respect to compact star binaries, leading to larger values of $(\Delta\alpha)^{2}$.

From the observed gravitational waveforms containing
at least one NS, it is possible to put independent
constraints on $\Delta \alpha$. This gives further
insights on the possible deviation from GR in
extreme gravity regimes.

\section{Horndeski GW forecasts}
\label{foresec}

\subsection{Degrees of freedom and approximations}

Even though here we are only considering scalar-tensor corrections up to 0.5PN, to perform accurate forecasts it is important to include higher-order PN terms even in their form predicted by GR, technically breaking consistency. The two most important reasons are: they make it possible to break the degeneracy between $\kappa_s (\Delta\alpha)^2$ and $\widetilde{m}$ by allowing a measure of the symmetric mass ratio $\eta$, and it is well known that higher-order PN terms degrade the chirp-mass constraints when including terms of order 1PN or higher, by a factor $\sim 3$~\cite{Arun:2004hn} if spins can be neglected, or larger than 10 if they cannot~\cite{Cutler:1994ys}. We thus rewrite the waveform including terms up to order 1.5PN for the fundamental harmonic and explicitly expanding the total redshifted mass $\widetilde{m}$ as
\begin{equation}
    \widetilde{m} = \frac{\widetilde{\cal M}}{\eta^{3/5}}\,,
\end{equation}
since $\widetilde{\cal M}$ and $\eta$ will be used as free parameters in our Fisher matrix instead of $\widetilde m$~\cite{Cutler:1994ys,Arun:2006yw,Mishra:2010tp}. We note that going to 1PN (2PN) or above in the amplitude (phase) have negligible impact on the uncertainties of parameters here considered.
We therefore write, absorbing henceforth $G_*$
into $\widetilde{\cal M}$\,:
\begin{align} \label{eq:hspa}
    \tilde{h}_c (f_0)
    = & \;\sqrt{\frac{5\pi}{24}} \frac{ \widetilde{\cal M}^{5/6}}
    {d_{\rm GW}(z)}  \left( \pi f_0 \right)^{-7/6}
    \left[ 1-\frac{5 \kappa_s (\Delta{\alpha})^2 \eta^{2/5}}{48\big(\pi \widetilde{\cal M} f_0 \big)^{2/3}}
    \right] e^{i \Psi_{+}}, \\
    \Psi_{+} = & \; 2\pi t_{c}f_{0}-2\Phi_{c}-\frac{\pi}{4}
    +  \frac{3}{128 \eta} \Bigg[
    \alpha_{-2} \kappa_s (\Delta{\alpha})^2 \bigg(\frac{\widetilde{\cal M}}{\eta^{3/5}}\pi f_{0} \bigg)^{-7/3}
    + \eta \Big(\widetilde{\cal M}\pi f_{0}\Big)^{-5/3}\nonumber \\
    & \;
    + \alpha_1\kappa_s (\Delta{\alpha})^2
    \Big(\frac{\widetilde{\cal M}}{\eta^{3/5}}\pi f_{0}\Big)^{-4/3}
    +\alpha_2(\eta)  \bigg(\frac{\widetilde{\cal M}}{\eta^{3/5}}\pi f_{0} \bigg)^{-3/3} +  \alpha_3(\beta) \bigg(\frac{\widetilde{\cal M}}{\eta^{3/5}}\pi f_{0} \bigg)^{-2/3}
    \Bigg] \,,
\end{align}
where $\alpha_{-2} = - 5/42$,
$\alpha_1= 25\pi/42$~\cite{Lang:2014osa,Bernard:2022noq}, $\alpha_2(\eta) \equiv 3715/756 + 55\eta/9$, and $\alpha_3 \equiv - 16\pi + 4\beta$~\cite{Cutler:1994ys}. Note that beside the leading-order ($-$1PN) extra scalar term in the phase, parameterized by $\alpha_{-2}$, we included the relative 1.5PN one, parameterized by $\alpha_1$, which is due to scalar waves scattered by the gravitational field of the binary, or tail effect, as GR does not give any contribution at this order, which is 0.5PN with respect to leading GR quadrupolar emission. For the 0PN, 1PN, 1.5PN terms we used the GR predicted values of the phasing coefficients, where, in particular, $\beta$ is the spin-orbit parameter, related to the individual dimension-less spins $\chi_1$ and $\chi_2$ through
\begin{equation}\label{eq:beta}
    \beta = \frac{113}{12} \frac{m_1-m_2}{m_1+m_2} \frac{\chi_1-\chi_2}{2}+ \left[ \frac{113}{12}-\frac{19}{3} \frac{m_1 m_2}{(m_1+m_2)^2} \right] \frac{\chi_1+\chi_2}{2} \,.
\end{equation}
We neglect the spin effect on the amplitude, and note that at 2PN there would be an extra spin-spin parameter.

We will assume for simplicity that the sky location of the source is always known exactly even for events where an electromagnetic counterpart would be hard to detect. The reason for this {is that GW FM inversion} is often a delicate issue, and having less parameters lead to more robust inversions. And in practice relaxing this assumption does not affect significantly the precision on the relevant parameters here considered.


The phase and amplitude parameters, $\theta_p$ and $\theta_a$, are therefore
\begin{equation}
    \theta_p=\big\{\Phi_{c},t_{c},\widetilde{\cal M},\eta, \beta,\kappa_s(\Delta{\alpha})^2 \big\}, \qquad
    \theta_a=\big\{\widetilde{\cal M},d_{\rm GW},\kappa_s(\Delta{\alpha})^2, \upsilon\equiv \cos \iota ,\psi \big\}\,.
\end{equation}
Overall, we can determine therefore the parameter
set\footnote{We remind the reader that we work at fixed sky localization.}
\begin{equation}
    \theta_t=\big\{\Phi_{c},t_{c}, \ln\widetilde{\cal M}, \eta, \beta,\ln d_{\rm GW}, \kappa_s(\Delta{\alpha})^2, \upsilon, \psi  \big\}\,,
\end{equation}
where we use $(\Delta{\alpha})^2$ instead of $\Delta{\alpha}$ or its logarithm to avoid singularities as our fiducial value is $\Delta{\alpha}=0$. The fiducials for the other parameters are discussed below in the next sub-section. If one combines the constraints on $d_{\rm GW}$ with those on the luminosity distance $d_L$ from, e.g., supernovae, then one can constrain also the effective Planck mass ratio $F(\phi_0)/F(\phi_s)$  [see, e.g., Eq.~(\ref{dua})], but we will not pursue this possibility here.

Let us now discuss in more detail how the inclination angle $\iota$ and the polarization angle $\psi$ affect the detector amplitude, as well as detector geometry parameters. We find convenient to
adopt the parametrization introduced in Ref.~\cite{Cutler:1994ys}  (see also Refs.~\cite{Chassande-Mottin:2019nnz,deSouza:2023gjv}), which expresses the detector polarization (\ref{eq:hdet}) as
\begin{equation}
\label{cut_par}
    \tilde{h}_a(f)= e^{- 2\pi i f \tau_a} \mathcal{A}_B
    \hat{F}^B_a(\hat{n})\,\tilde{k}\big(f;t_c, \mathcal{M},\eta,\Phi_c, \beta, d_{\rm GW},\kappa_s(\Delta\alpha)^2\big)\,,
\end{equation}
where capital Latin indices stand for polarization indices $\{+,\times\}$, small Latin ones for detector indices, and $\hat{F}_a^A$ is the antenna pattern of detector $a$ for polarization $A$.  {With $\tilde k$ we denoted the function returning the frequency dependence of the waveform, which does not depend on parameters involving the relative orientation between source and observer, that instead are encoded in the pattern function $\hat F$ and in $\mathcal{A}.$}
The inclination-dependent part has been re-arranged in a 2-vector $\chi_A$ defined as
\begin{equation}\label{eq:chi}
    \chi_+(\upsilon) \equiv \dfrac{1+\upsilon^2}{2},
    \qquad
    \chi_\times(\upsilon)\equiv -i\upsilon\,,
\end{equation}
which have been absorbed into the newly defined 2-vector ${\cal A}_A$
\begin{equation}
\label{eq:defa}
    \mathcal{A}_A(\upsilon,\psi)\equiv
    R^B_A (2 \psi) \chi_B (\upsilon)\,.
\end{equation}

Detector output $h_a$ is processed through matched-filtering, which can be expressed as a scalar product between $h_a$ and a template waveform $h_t$ in the space of functions \cite{mf,LIGOScientific:2019hgc}
\be
\label{sc_prod}
\langle h_a|h_t\rangle=4 \, \Re\int_0^\infty
\frac{h_a h_t^*}{S_{na}} \dd f\,,
\ee
whose integral is weighted by the spectral noise $S_{na}$ of the detector. From the scalar product (\ref{sc_prod}) one can define a norm of the gravitational waveform,  which in the parametrization (\ref{cut_par}), can be written as
\begin{equation}
\label{eq:scal_p_sec3}
|h|^2=\braket{h|h} =\Re\left[\mathcal{A}^*_A
\mathcal{A}_B\Theta^{AB}\right]
\braket{\tilde{k}|\tilde{k}}\,,
\end{equation}
where the matrix $\Theta^{AB}$, which collects the contributions of several detectors, is defined by
\begin{equation}
    \Theta^{AB}(\hat{n})\equiv
    \hat{F}_a^A(\hat{n})
    \hat{F}_b^B(\hat{n}) \delta^{ab} \dfrac{\int_{0}^{\infty}|\tilde k(f)|^2 S_{n,a}^{-1} \mathrm{d}f}{\int_0^{\infty}|\tilde k(f)|^2S_{n,{\rm aver}}^{-1}(f)\mathrm{d}f}\,.
\end{equation}
In the above $S_{n,{\rm aver}}$ is the average over all detectors of the spectral noise densities:
\begin{equation}
    S_{n,{\rm aver}}^{-1} (f)\equiv n_d^{-1}
    \sum_{a=1}^{n_d} S_{n,a}^{-1}(f)\,.
\end{equation}
In GR one has $|\tilde k(f)|^2\propto f^{-7/3}$. In Horndeski theories this is no longer exactly the case because of Eq.~\eqref{eq:hspa}, but here we will assume that this is still a good approximation since $\kappa_s (\Delta{\alpha})^2$ is expected to be small.  We note that in writing the above equations we implicitly decomposed the full polarization angle into a term $\psi$ which only depends on the source, and absorbed the detector-dependent part into the pattern functions $\hat F_a^A(\hat n)$ (see~\cite{deSouza:2023gjv} for a detailed discussion).

The general Fisher Matrix (FM) for the full set of parameters in detector $a$ is given by
\begin{align} \label{eq:FMtot}
    {\cal F}_{a,\alpha\beta} &=4 \,\Re\Bigg[\int_{0}^{\infty}\frac{h_{ta,\alpha}h_{ta,\beta}^{*}}{S_{na}(f)} \dd f \Bigg]
    \nonumber\\
    & = \Re\Bigg[
    \hat{F}^A_a  \hat{F}^B_a \bigg(
     {\cal A}_{A,\alpha}{\cal A}^*_{B,\beta}\braket{\tilde{k}|\tilde{k}}
    +{\cal A}_{A,\alpha}{\cal A}^*_{B}\braket{\tilde{k}|\tilde{k}_{,\beta}}
    +{\cal A}_{A}{\cal A}^*_{B,\beta}\braket{\tilde{k}_{,\alpha}|\tilde{k}}    +{\cal A}_{A}{\cal A}^*_{B}\braket{\tilde{k}_{,\alpha}|\tilde{k}_{,\beta}}
    \bigg)_{S_{na}} \Bigg]\,.
\end{align}
Note that derivatives do not act on the terms involving the pattern functions. We also have defined
\begin{equation}
    \braket{\tilde{k}|\tilde{k}'}_S
    =4\int_{0}^{\infty} \dfrac{\tilde{k}(f)\tilde{k}^{'*}(f)}
    {S(f)} \mathrm{d}f\,,
\end{equation}
and to lighten notation we have moved the suffix $S_{na}$ common to all scalar products outside the parenthesis in Eq.~(\ref{eq:FMtot}).

Following Ref.~\cite{Cutler:1994ys}, we further note that the symmetric matrix $\Theta^{AB}$ can be diagonalized by a rotation $R_{AB}(2\bar\psi)$, i.e., by re-defining effective polarizations $\bar{+}$ and $\bar{\times}$. The resulting diagonal $\bar{\Theta}^{AB}$ to $\bar{+}$ and $\bar{\times}$ takes the simple form
\begin{equation}
\bar{\Theta}^{AB}
= \sigma_d \left(\begin{array}{cc}
1+\epsilon_d & 0\\
0 & 1-\epsilon_d
\end{array}\right),
\label{eq:thetadiag}
\end{equation}
where $\sigma_d \equiv (\Theta^{++} + \Theta^{\times \times})/2$ and $0\leq \epsilon_d \leq 1$. Both $\sigma_d$ and $\epsilon_d$ depend only on the sky location of the source, and can be computed in terms of the antenna patterns $\hat{F}^A_a(\hat{n})$ for a given detector.
We define the  total network signal-to-noise ratio $\rho$ of a given event as
\begin{equation}
    \rho = \sqrt{\bar{\Theta}^{AB} \bar{\cal A}_A\bar{\cal A}^*_B\braket{\tilde{k}|\tilde{k}}_{S_{n\rm avg}}}\,.
\end{equation}
When $\epsilon_d$ is close to unity, the detector is unable to measure well both polarization amplitudes, which leads to degenerate posteriors for $d_{\rm GW}$, $\upsilon$ and $\psi$. In practice, this degeneracy appears when $\epsilon_d \gtrsim 1-1/\rho$~\cite{Chassande-Mottin:2019nnz}. In order to preserve our FM inversion accuracy, we therefore discard all events for which $\epsilon_d > 0.95$.

The final full FM, summing over all detectors, can be written as
\begin{equation}
    {\cal F}_{\alpha \beta} = \sum_a {\cal F}_{a,\alpha \beta} = \bar{\Theta}^{AB} \, \Re \bigg( \bar{\cal A}_{A,\alpha}\bar{\cal A}^*_{B,\beta}\braket{\tilde{k}|\tilde{k}}
    +\bar{\cal A}_{A,\alpha}\bar{\cal A}^*_{B}\braket{\tilde{k}|\tilde{k}_{,\beta}}
    +\bar{\cal A}_{A}\bar{\cal A}^*_{B,\beta}\braket{\tilde{k}_{,\alpha}|\tilde{k}}
    +\bar{\cal A}_{A}{\cal A}^*_{B}\braket{\tilde{k}_{,\alpha}|\tilde{k}^{*}_{,\beta}}\bigg)_{S_{n\rm avg}}\,,
\end{equation}
where $\bar {\cal A}_A \equiv R_A^B\big(2(\bar\psi + \psi)\big) \chi_B(\upsilon)$. We remark that $\bar\psi$ is completely determined by sky and detector localization, so in a FM analysis like ours the only thing that matters in terms of polarization angle is that $\psi$ is the only  degree of freedom, which is drawn for each event from an uniform distribution. We will thus no longer mention $\bar\psi$ in what follows.

Note that, anytime the terms inside the parenthesis in Eq.~\eqref{eq:FMtot} are imaginary, it will result in a null entry ${\cal F}_{\alpha \beta}$. This is the case for the $\{\ln d_{\rm GW},\,\Phi_c\}$, $\{\ln d_{\rm GW},\,t_c\}$, $\{\ln d_{\rm GW},\,\beta\}$ and $\{\ln d_{\rm GW},\,\eta\}$ terms.

\subsection{Forecasts for Advanced LIGO and Einstein Telescope}

We are now ready to make forecasts for both BNS and NSBH mergers for two observational settings, Advanced LIGO (aLIGO) and the Einstein Telescope (ET). As discussed above we keep terms up to 1.5PN, after which the precision on the parameters affected by the non-GR terms converge. That is, the new parameters appearing in 2PN and above should have similar constraints as in GR. Since NSBH mergers can also produce detectable EM counterparts~\cite{Barbieri:2019kli} (although perhaps only a fraction of them may do so~\cite{Fragione:2021cvv}), we assume here for simplicity that such counterpart would be detected such as to fix the position of events in the sky precisely. Nevertheless, for our results, the so-called extrinsic GW parameters (angles, distances and inclinations) play a very limited role as we find that 99\% of the constraining power on the $-1$PN parameter comes from the phase, not the amplitude. We also remark that in practice the 0.5PN term makes little difference for our forecasts, to wit within $2\%$ in all parameters considered.

For Advanced LIGO we assume the design sensitivity~\cite{TheLIGOScientific:2014jea}, which is slightly better than what was achieved so far in O1--O3. In particular, we assume a minimum frequency of 11 Hz. This contrasts to the cut-off at 30 Hz imposed for instance in Ref.~\cite{LIGOScientific:2021sio}. For the Einstein Telescope, we assume the complete 3-detector triangular configuration with sensitivities described by the so-called ET-D curve~\cite{Hild:2010id}, and take 2 Hz as the minimum achievable frequency. We note that, for the binary mergers here considered, the assumed minimum frequency plays an important role in the final precision.

The fundamental event rates for both BNS and NSBH coalescences are still very uncertain. Therefore, instead of forecasting what a given observing run may be able to constrain collectively, we focus on forecasts for the expected capability of a single event. Since these depend on the intrinsic and extrinsic properties, we make use of expected mass, spin, and distance distributions, and depict results obtained by randomly selecting 10000 events. Each event has a random sky position and inclination parameter $\upsilon$ assuming uniform distributions. We also assume a comoving rate of events constant in redshift, so that they are uniformly distributed in the sky volume. For this we assume as fiducial a flat $\Lambda$CDM cosmology with the density  parameter of non-relativistic matter $\Omega_{m0}=0.3$ and $H_0 = 70$km/s/Mpc. This means that distance distribution is an increasing function of $d_{\rm GW}$.

We note that forecasts combining a given number of events could be obtained trivially for universal parameters, but all parameters here considered  are in principle unique for each event. So a forecast of how many events would be needed in order to rule-out modifications to GR with GWs would only be possible if one included a prior knowledge on the distribution of $\kappa_s(\Delta{\alpha})^2$. Since this is model-dependent, we do not perform this analysis here.

Following Ref.~\cite{Iacovelli:2022bbs}, we assume that NS masses are uniformly distributed in the range $[1, 2.5]M_{\odot}$ and that BH masses are distributed according to Eq.~(6) of Ref.~\cite{Zhu:2020ffa}. Since for Advanced LIGO the resulting FM uncertainty in $\beta$ is often much larger than what is allowed in GR, we add a prior to the parameters $\chi_1$ and $\chi_2$ and compute the corresponding prior on $\beta$ as the standard deviation obtained through Eq.~\eqref{eq:beta}. We consider two possible priors for $\chi_{1,2}$. The standard prior, which assumes $\chi_{\rm NS} \in {\cal U}[-0.05, 0.05]$ and $\chi_{\rm BH} \in {\cal N}(0, 0.15)$, is what we consider in the main text, where ${\cal N}$ and ${\cal U}$ are the Gaussian  and uniform distributions, respectively. In Appendix~\ref{app:prior} we also show results for a minimal prior corresponding to $\chi_{\rm NS}, \chi_{\rm BH} \in {\cal U}[-1, 1]$.  For the polarization angle we use $\psi\in {\cal U}[0, \pi]$. We also set $t_c = \Phi_c = 0$ as fiducials for all events as we are not interested in these quantities here. Finally, we discard all events for which the total network signal to noise ratio $\rho$  is smaller than 8, but we also show results including only high S/N events, with $\rho \ge 30$.

\begin{figure}[t]
\centering
\includegraphics[width=0.45\textwidth]{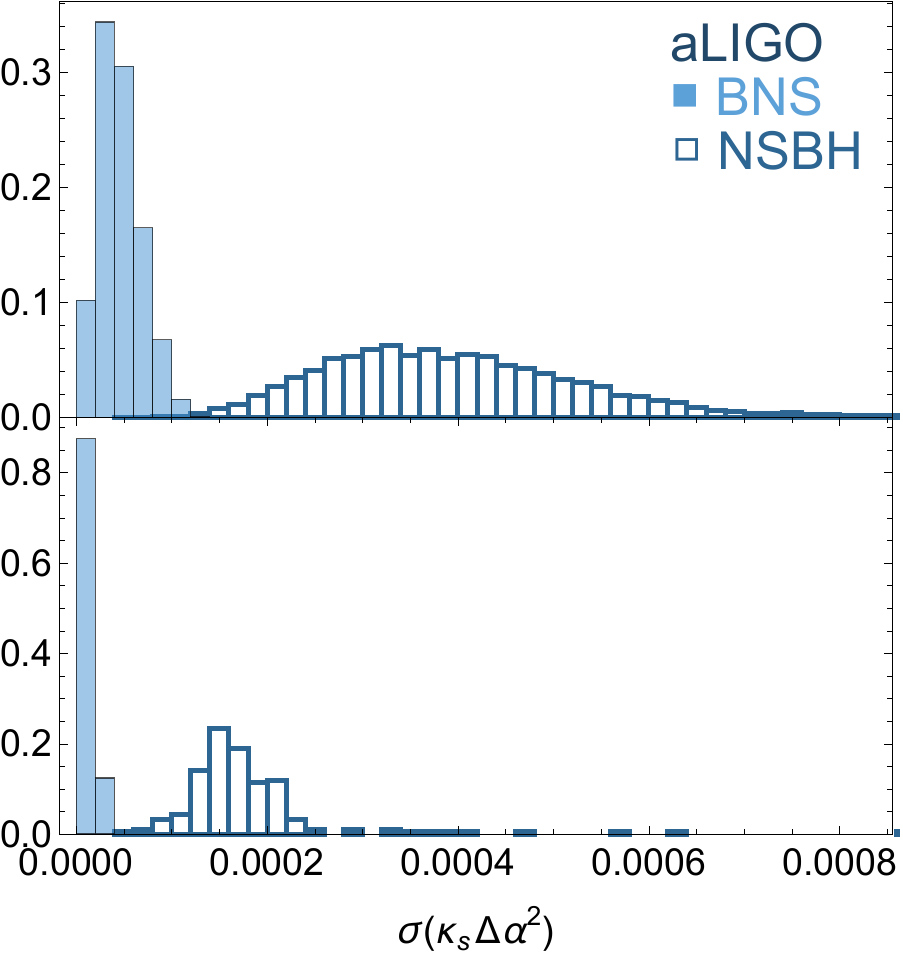}\quad\;
\includegraphics[width=0.45\textwidth]{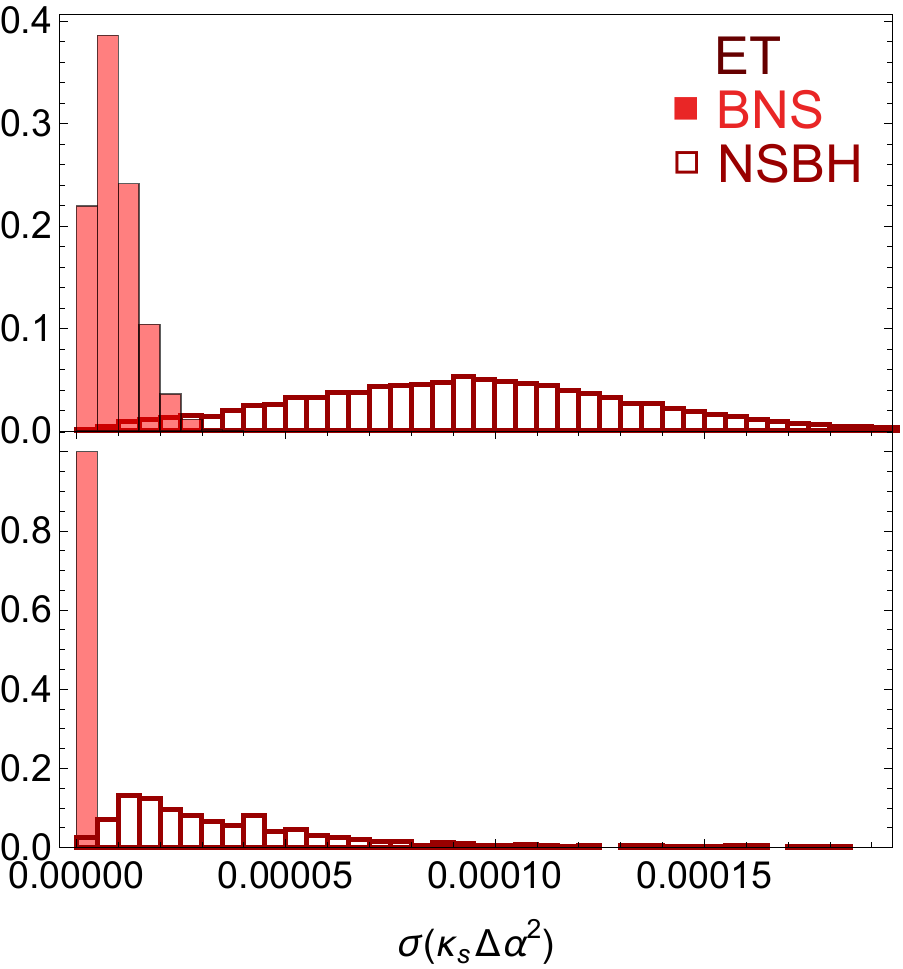}
\caption{
    Distribution of forecast errors on $\kappa_s(\Delta{\alpha})^2$ for a single GW event with EM counterpart measured by assuming standard spin priors.  Solid (open) histograms are for BNS (NSBH) mergers. \emph{Left:} Advanced LIGO. \emph{Right:} Einstein Telescope. \emph{Top:} all cases that the network S/N $\rho \ge 8$; \emph{Bottom:} only events with $\rho \ge 30$. BNS mergers have $\sim10$ ($\sim20$) times smaller uncertainties than NSBH ones for $\rho \ge 8$ ($\rho \ge 30$).  We include all terms up to 1.5PN in the phase.
    }\label{fig:forecasts-alpha}
\end{figure}

\begin{figure}[t]
    \centering
    \includegraphics[width=0.45\textwidth]{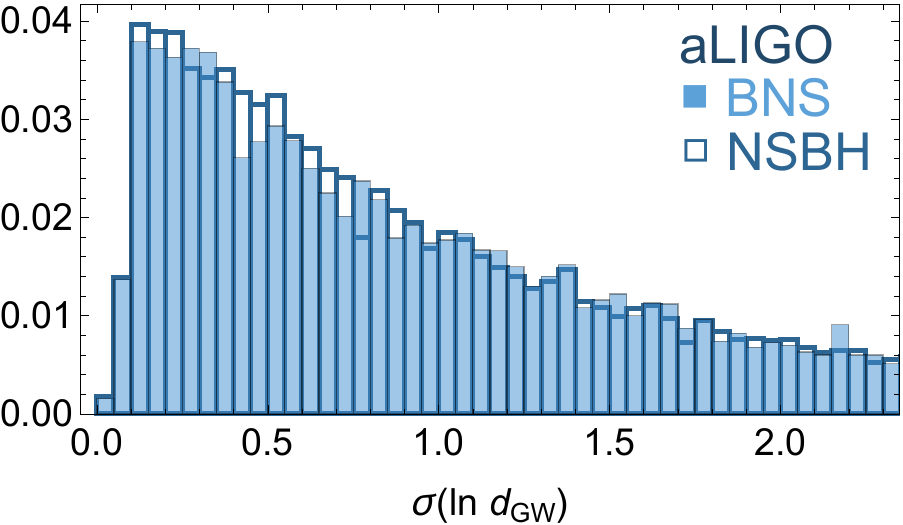}\quad
    \includegraphics[width=0.45\textwidth]{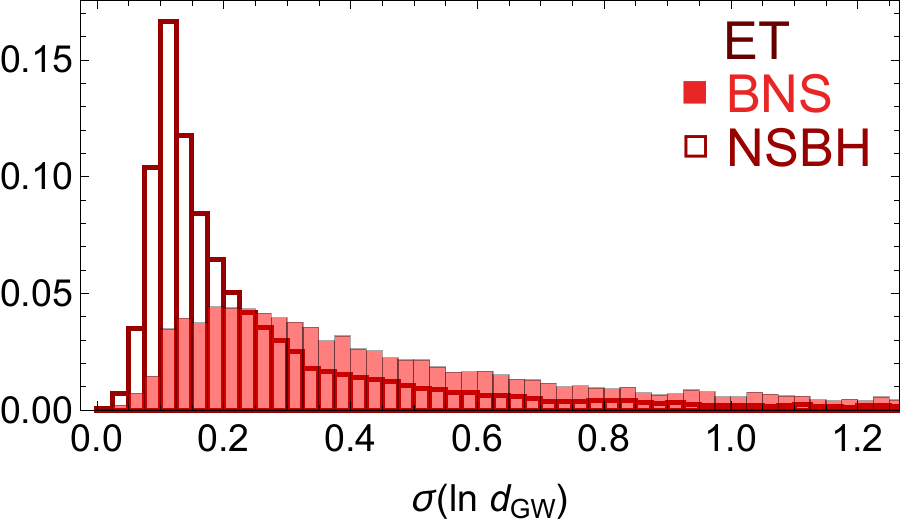}
    \includegraphics[width=0.45\textwidth]{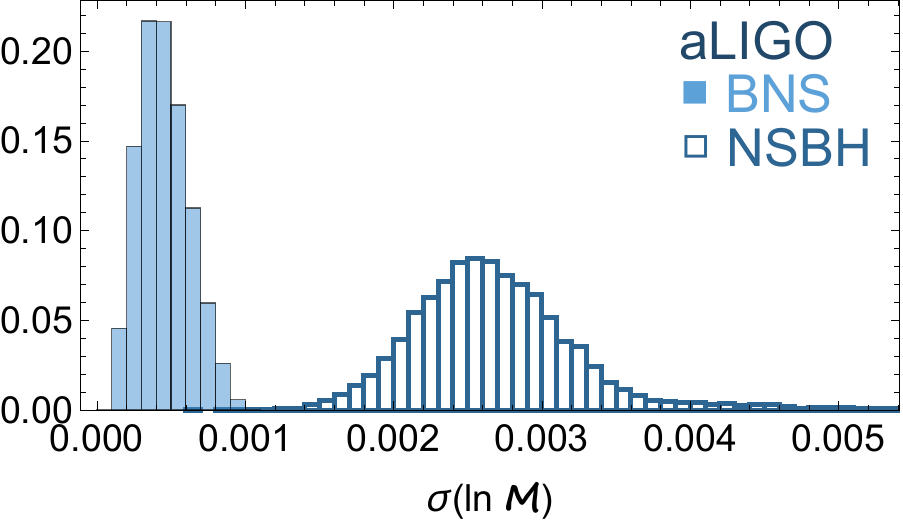}\quad
    \includegraphics[width=0.45\textwidth]{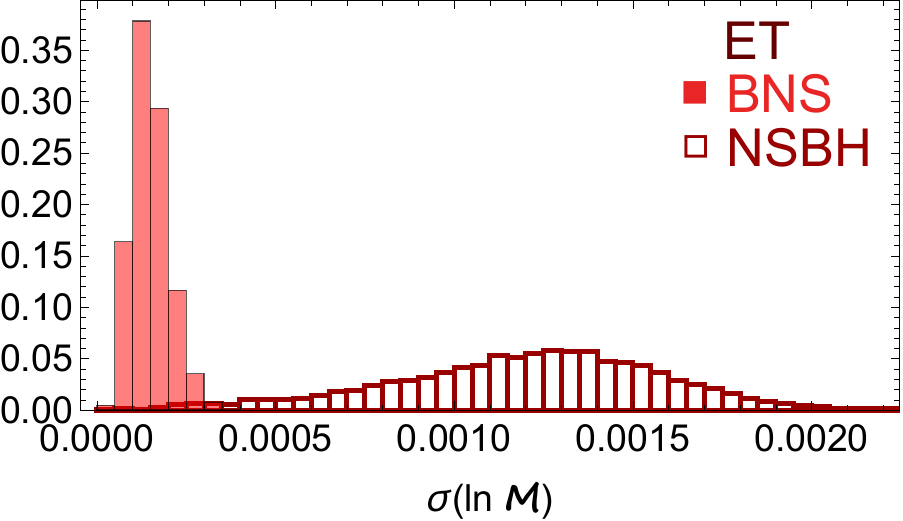}
    \includegraphics[width=0.445\textwidth]{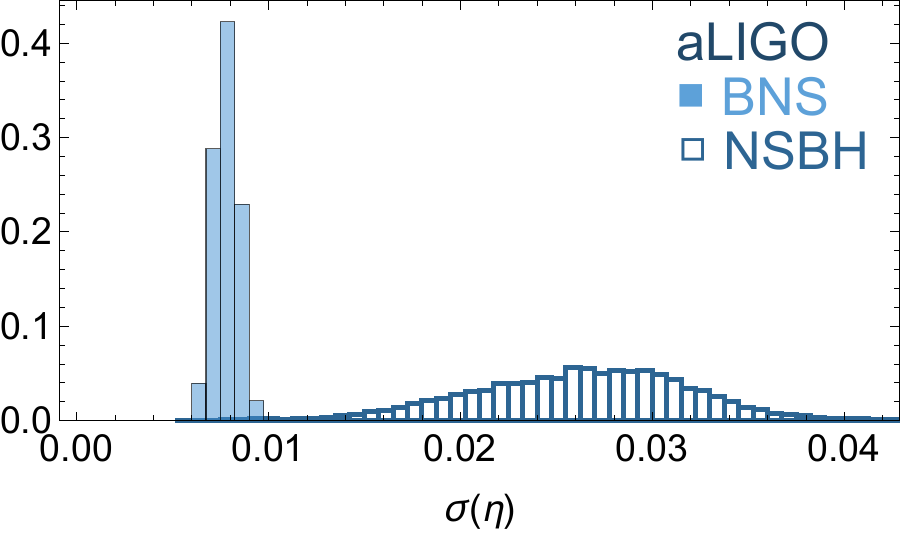}\quad\;
    \includegraphics[width=0.445\textwidth]{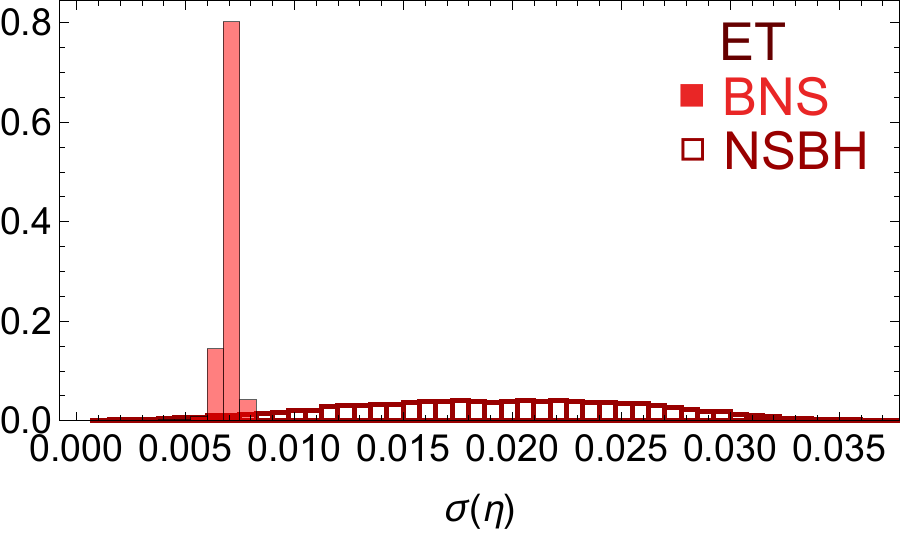}
    \caption{Same as Figure~\ref{fig:forecasts-alpha} but for the distance to the merger $d_{\rm GW}$, chirp mass ${\cal M}$,
    and symmetric mass ratio $\eta$, for all events with SNR $\rho \ge 8$. BNS events have around 7 times better precision than NSBH ones for ${\cal M}$ and 3 times better for $\eta$.
    }\label{fig:forecasts-extra}
\end{figure}

\begin{figure}[t]
    \centering
    \includegraphics[width=0.45\textwidth]{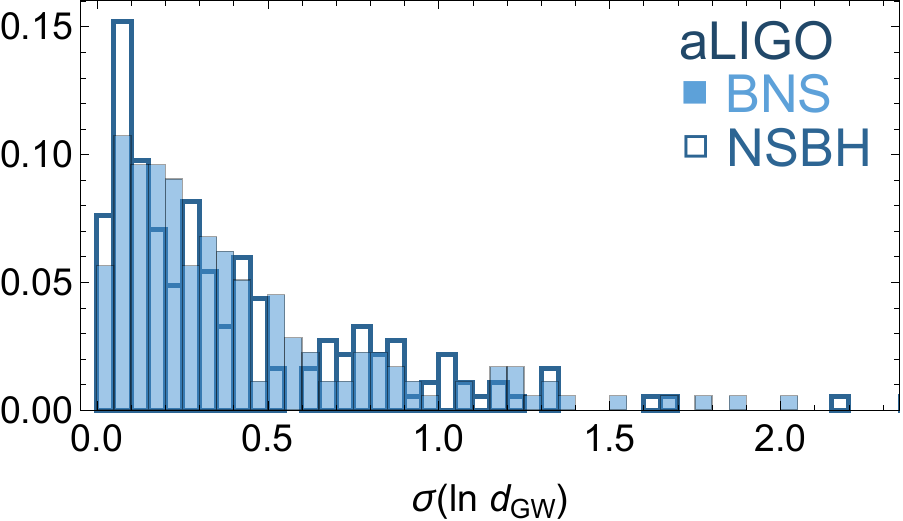}\quad
    \includegraphics[width=0.45\textwidth]{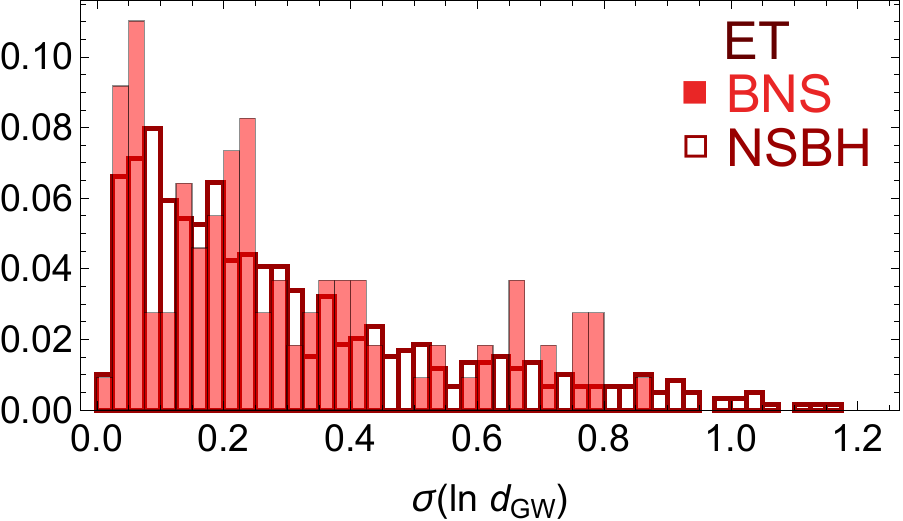}
    \includegraphics[width=0.45\textwidth]{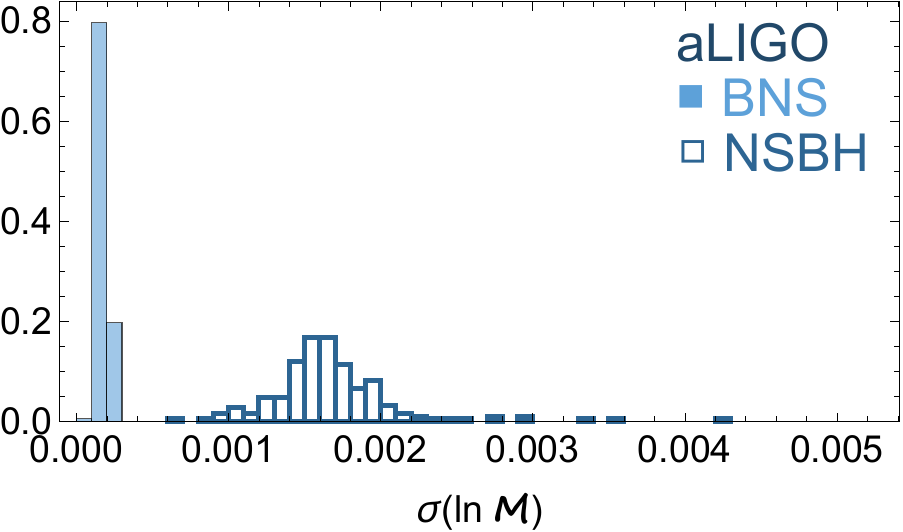}\quad
    \includegraphics[width=0.45\textwidth]{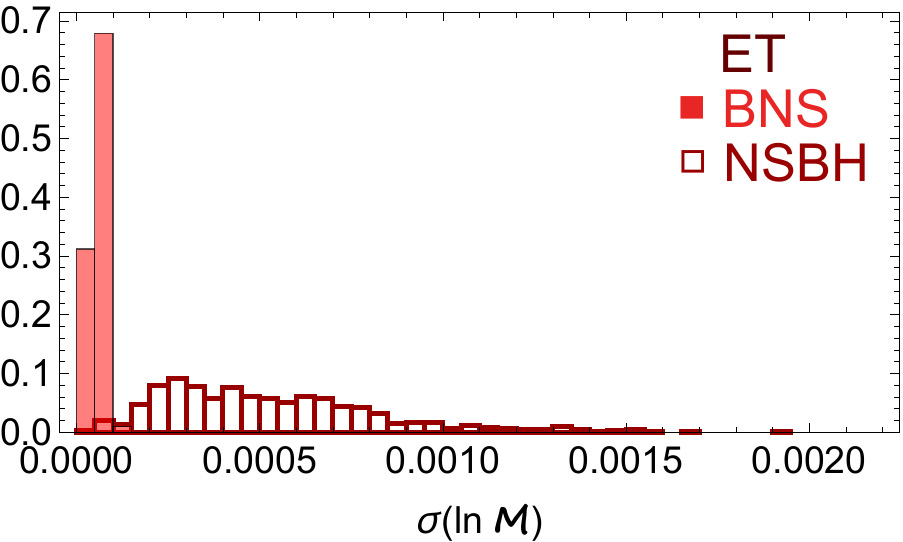}
    \includegraphics[width=0.45\textwidth]{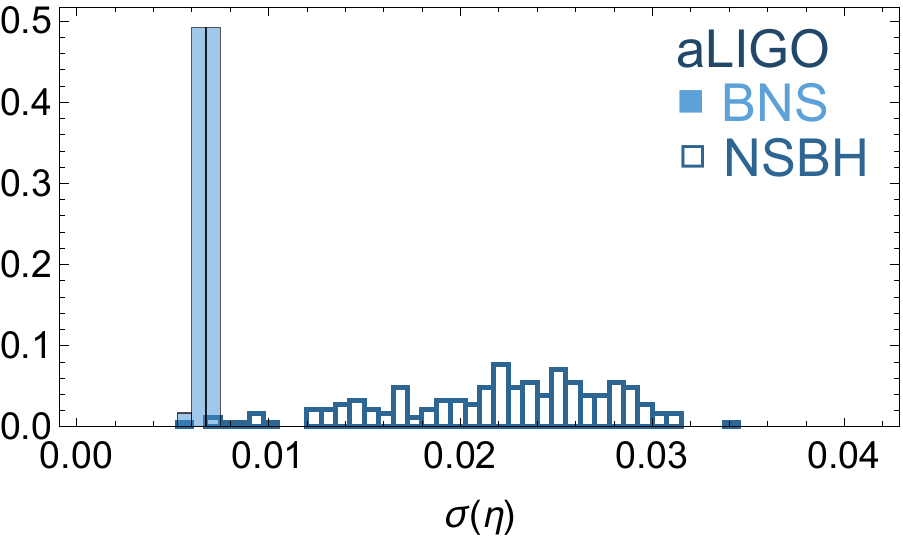}\quad
    \includegraphics[width=0.45\textwidth]{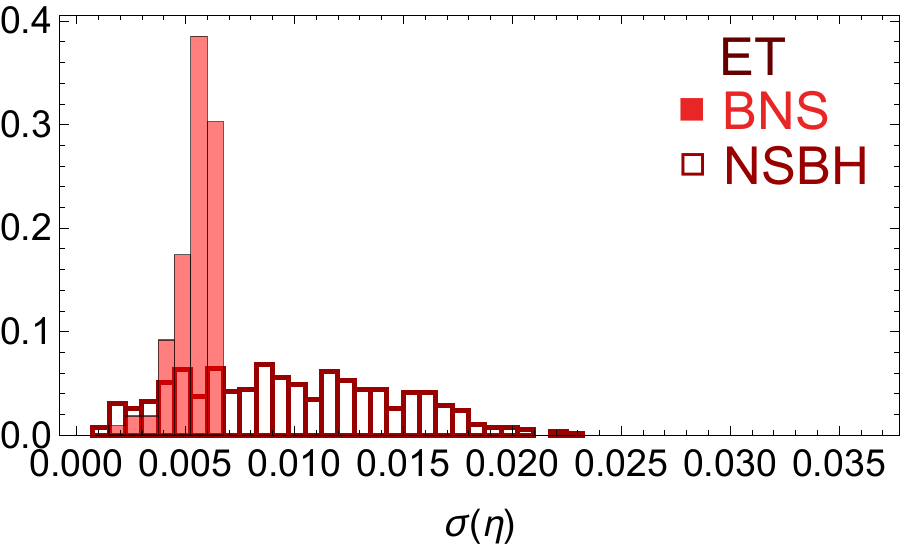}
    \caption{Same as Figure~\ref{fig:forecasts-extra} but only for events with SNR $\rho \ge 30$. BNS events have even better relative precision in ${\cal M}$ than NSBH ones, while for $\eta$ relative results are similar and for $d_{\rm GW}$ both BNS and NSBH become equivalent.
    }\label{fig:forecasts-SNR30}
\end{figure}

Figure~\ref{fig:forecasts-alpha} shows the distribution of marginalized constraints in the parameter $\kappa_s (\Delta{\alpha})^2$ for both aLIGO and ET sensitivities, for both BNS and NSBH mergers. We depict both all events with total network S/N $\rho \ge 8$ and only the high significance events with $\rho \ge 30$. Figure~\ref{fig:forecasts-extra} shows similar histograms for the parameters  $d_{\rm GW}$, $\ln {\cal M}$ and $\eta$ for all events with $\rho \ge 8$. In Figure~\ref{fig:forecasts-SNR30}, we illustrate instead these parameters for the subgroup with $\rho \ge 30$. These are drawn from the 10000 Monte Carlo samples in each case, and since the number of events fall sharply with increasing $\rho$, the number of GWs in each of these plots are only few hundreds. We note that BNS mergers can give constraints which are much tighter than those from NSBH. We will compare the perspectives measuring a scalar charge with either BNS and NSBH below.


\begin{table}
\centering
    \begin{tabular}{c|ccccccccc}
    & $\Phi_{c}$ & $t_{c}$ & $\ln\widetilde{\cal M}$ & $\eta$ & $\beta$ & $\ln d_{\rm GW}$ & $\kappa_s(\Delta{\alpha})^2$ & $\upsilon$ & $\psi$ \\ \hline
    $\Phi_{c}$ &  1 & -0.636 & 0.579 & -0.915 & -0.885 & 0 & 0.417 & $(\pm 0.20)$ & $(\pm 0.97)$ \\
    $t_{c}$ &  -0.636 & 1 & -0.128 & 0.377 & 0.343 & 0 & -0.070 & $(\pm 0.13)$ & $(\pm 0.62)$ \\
    $\ln\widetilde{\cal M}$ & 0.579 & -0.128 & 1 & -0.817 & -0.859 & 0 & 0.963 & $(\pm 0.11)$ & $(\pm 0.56)$ \\
    $\eta$ & -0.915 & 0.377 & -0.817 & 1 & 0.997 & 0 & -0.658 & $(\pm 0.18)$ & $(\pm 0.89)$ \\
    $\beta$ & -0.885 & 0.343 & -0.859 & 0.997 & 1 & 0 & -0.710 & $(\pm 0.18)$ & $(\pm 0.86)$ \\
    $\ln d_{\rm GW}$ & 0 & 0 & 0 & 0 & 0 & 1 & 0 & $(\pm 0.95)$ & $(\pm 0.15)$  \\
    $\kappa_s(\Delta{\alpha})^2$ & 0.417 & -0.070 & 0.963 & -0.658 & -0.710 & 0 & 1 & $(\pm 0.08)$ & $(\pm 0.41)$ \\
    $\upsilon$ & $(\pm 0.20)$ & $(\pm 0.13)$ & $(\pm 0.11)$ & $(\pm 0.18)$ & $(\pm 0.18)$ & $(\pm 0.95)$ & $(\pm 0.08)$ & 1 & $(\pm 0.08)$ \\
    $\psi$ & $(\pm 0.97)$ & $(\pm 0.62)$ & $(\pm 0.56)$ & $(\pm 0.89)$ & $(\pm 0.86)$ & $(\pm 0.15)$ & $(\pm 0.41)$ & $(\pm 0.08)$ & 1
    \end{tabular}
    \caption{Correlation between the Fisher Matrix parameters, obtained averaging over 10000 realizations.  Most of the correlation coefficients change very little in each realization, except the ones involving $\upsilon$ and $\psi$; for these variables the average correlation is zero, so we quote instead its standard deviation. Very similar results are obtained for ET and/or for NSBH mergers.}
    \label{tab:corr}
\end{table}

In Table~\ref{tab:corr}, we list the correlation matrix between the different parameters.  We represent the case of aLIGO and BNS averaging over 10000 realizations, but very similar results are obtained for ET and/or for NSBH mergers. Most of the correlation coefficients change very little from realization to realization, except the one between $d_{\rm GW}$ and $\upsilon$; this correlation varies over the whole possible range $[-1, 1]$, and thus although it tends to zero on average for any particular event, it can be very high. As can be seen, at 1.5PN, $\eta$ and $\beta$ are almost completely degenerate. Moreover, $\kappa_s (\Delta \alpha)^2$ is mostly correlated with ${\cal M}$, but is also with $\Phi_c$, $\eta$ and $\beta$.

We summarize the typical precision achievable in each BNS or NSBH merger in Table~\ref{tab:aver-sigma} for both aLIGO and ET. To wit, we show the average precision over the 10000 Monte Carlo realizations. We also show the ratio of precision in NSBH and BNS events. Note that BNS events result in better precision for all parameters except distance. In particular, $\kappa_s(\Delta{\alpha})^2$ is on average around 10--20 times more constrained in BNS than in NSBH. Nevertheless, the expected signal in NSBH systems is both higher and less degenerate, as only one scalar-tensor parameter $\hat{\alpha}_A$ is non-zero.

So for a given amount of possible scalar charge, which binary merger is a best tool to observe it? We explore this issue in detail in Appendix~\ref{app:Delta-alpha}, but the summary is that although between $\sim 15-40\%$ of the events will have a much suppressed scalar charge signal in BNS coalescences, in other cases the signal is still large. In particular, on average $(\Delta{\alpha})^2$ is expected to be typically around half of the maximum value of both $\alpha_A$ and $\alpha_B$, so that for a given modified gravity model with scalar charge, BNS mergers will have a higher chance than NSBH mergers of producing a significant falsification of GR.


It is interesting to compare the Horndeski forecasts to the GR ones. This is achieved in a straightforward manner in our FM by fixing the $\kappa_s(\Delta{\alpha})^2$ parameter, to wit by removing the corresponding row and column from the FM. Doing so we find that precision in $d_{\rm GW}$ is unchanged, which is expected as the dominant contribution of the scalar charge is on the --1PN term, not the amplitude, as discussed above. For the mass parameters, we see that  precision becomes considerably worse in Horndeski. In particular, for aLIGO, precision in $\widetilde{\cal M}$ degrades by factors of 5 (3) for BNS (NSBH), while ${\eta}$ only by 1.07 (1.2) for BNS (NSBH). For ET, $\widetilde{\cal M}$ degrades by factors of 3 (4) for BNS (NSBH), while ${\eta}$ by 1.05 (1.9) for BNS (NSBH). Moreover, we see that in the GR case results for $d_{\rm GW}$, $\widetilde{\cal M}$, and $\eta$ become broadly consistent with e.g.~\cite{Iacovelli:2022bbs}, which validates the accuracy of our code.

\setlength\tabcolsep{6pt}
\begin{table}
\centering
    \begin{tabular}{ccccccc}
    \hline
    \hline
    &\multicolumn{3}{c}{\rm \textbf{aLIGO}}&\multicolumn{3}{c}{\rm \textbf{ET}} \\ 
    &{\rm BNS}&{\rm NSBH}& $\frac{\rm NSBH}{\rm BNS}$ & {\rm BNS}&{\rm NSBH} & $\frac{\rm NSBH}{\rm BNS}$ \\
    \cline{2-7}
    & \multicolumn{6}{c}{$\rho \ge 8$}\\
    \cline{2-7}
    $\langle\sigma\ln\widetilde{\cal M}\rangle $      & 0.00049 & 0.0027 & 5.9 & 0.00015 & 0.0012 & 8.1  \\
    $\langle\sigma\eta\rangle $                       & 0.0078  & 0.026  & 3.4 & 0.0070 & 0.019   & 2.7   \\
    $\langle\sigma\beta\rangle $                      & 0.16    & 1.1    & 6.5 & 0.16   & 0.77    & 4.8   \\
    $\langle\sigma\ln d_{\rm GW}\rangle $             & 1.5     & 1.4   & 0.95 & 0.65   & 0.29    & 0.45   \\
    $\langle\sigma\kappa_s(\Delta{\alpha})^2\rangle$ & 0.000047& 0.00042 & 9.0 & $9.5\,10^{-6}$ & $9.7\,10^{-5}$ & 10  \\
    \cline{2-7}
    & \multicolumn{6}{c}{$\rho \ge 30$}\\
    \cline{2-7}
    $\langle\sigma\ln\widetilde{\cal M}\rangle $      & 0.00017 & 0.0017 & 10   & 0.000057 & 0.00053 &  9.4  \\
    $\langle\sigma\eta\rangle $                       & 0.0067 & 0.021   & 3.3  & 0.0055   & 0.0098  &  1.8  \\
    $\langle\sigma\beta\rangle $                      & 0.16 & 0.86      & 5.3  & 0.12    & 0.38     &  3.1  \\
    $\langle\sigma\ln d_{\rm GW}\rangle $             & 0.45 & 0.44      & 0.97  & 0.28    & 0.30     &  1.08 \\
    $\langle\sigma\kappa_s(\Delta{\alpha})^2\rangle$ & 0.000013& 0.00018 & 14 & $1.6\,10^{-6}$ & $3.8\,10^{-5}$ & 23  \\
    \hline
    \hline
    \end{tabular}
    \caption{Average precision (over 10000 simulated events) on each parameter from a single for BNS or NSBH coalescence.
    In columns 4 and 7, we also show the ratio of precision in NSBH and BNS events. We assume the standard spin priors,
    see the main text for detail.}
\label{tab:aver-sigma}
\end{table}

\begin{figure}[t]
    \centering
    \includegraphics[width=0.49\textwidth]{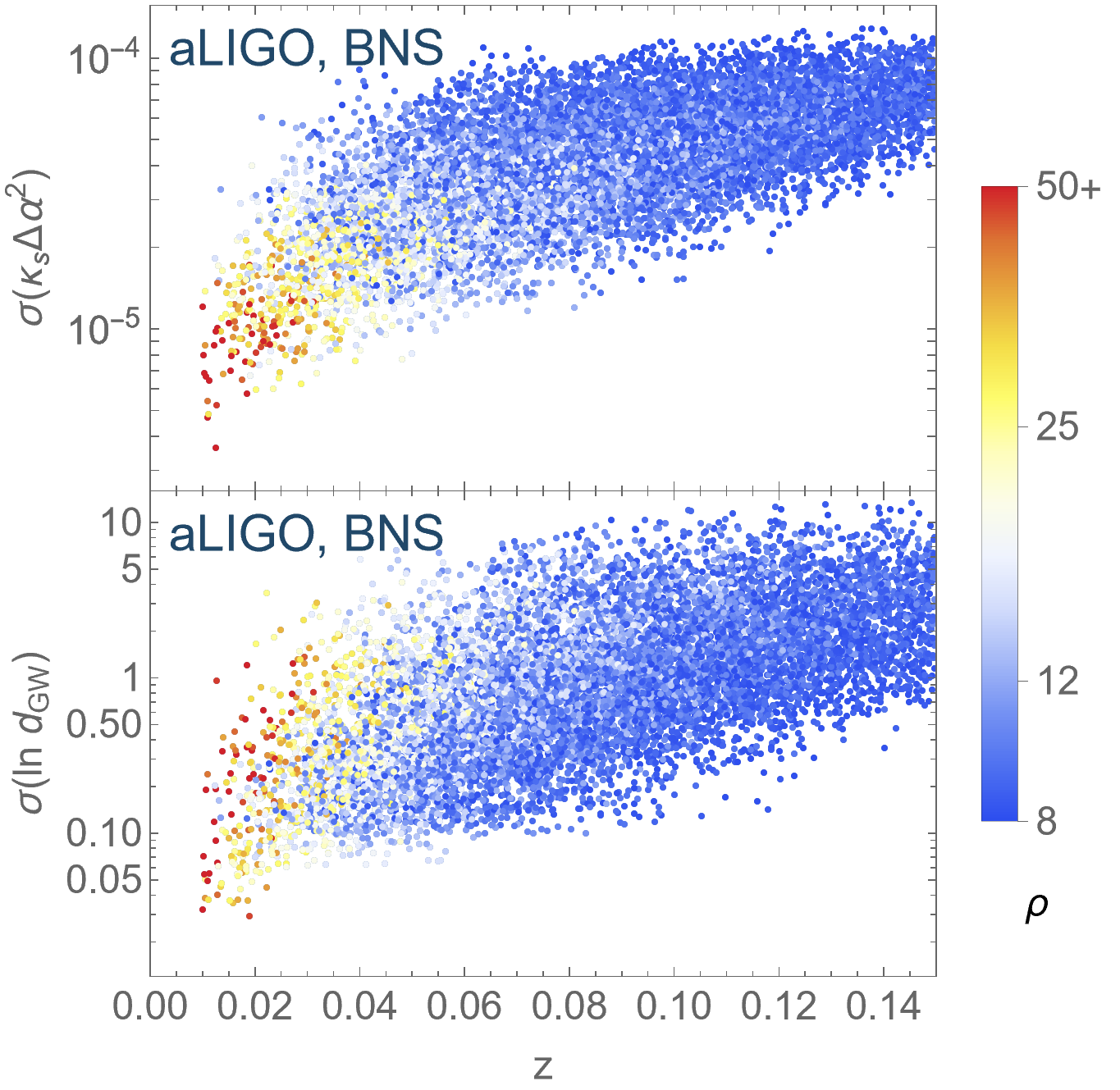}\quad
    \includegraphics[width=0.49\textwidth]{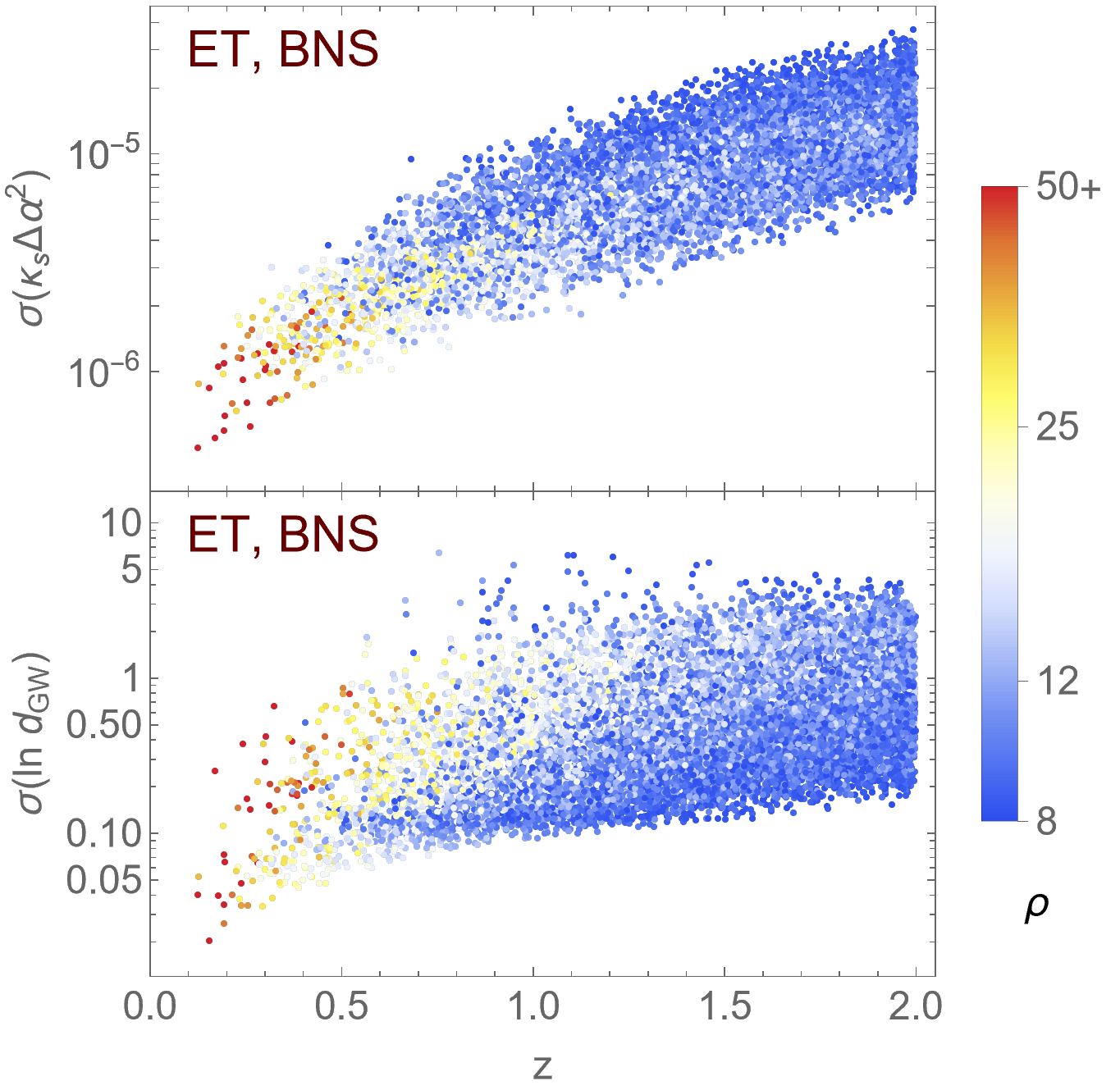}
    \vspace{0.1cm}
    \includegraphics[width=0.49\textwidth]{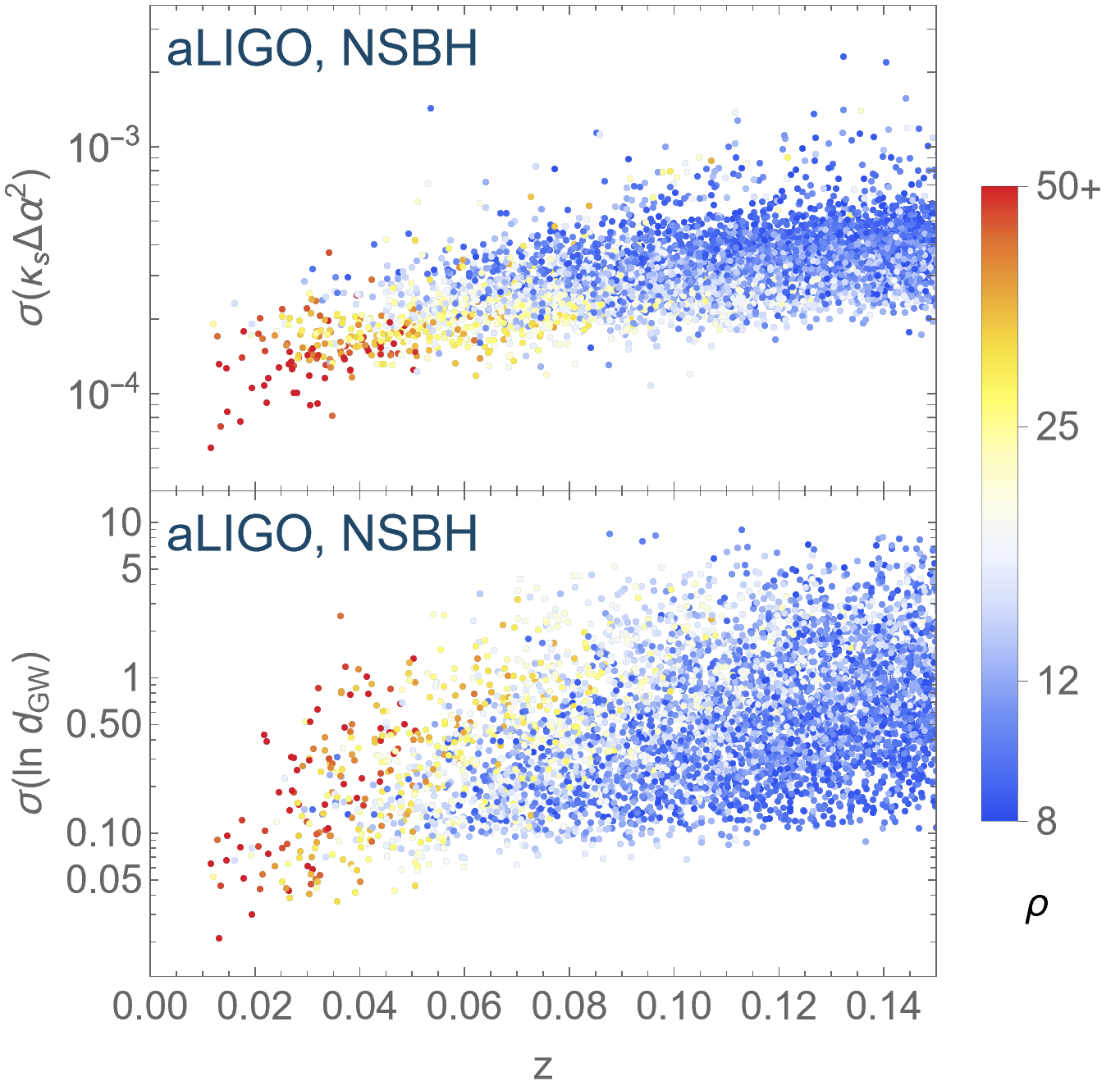}\quad
    \includegraphics[width=0.49\textwidth]{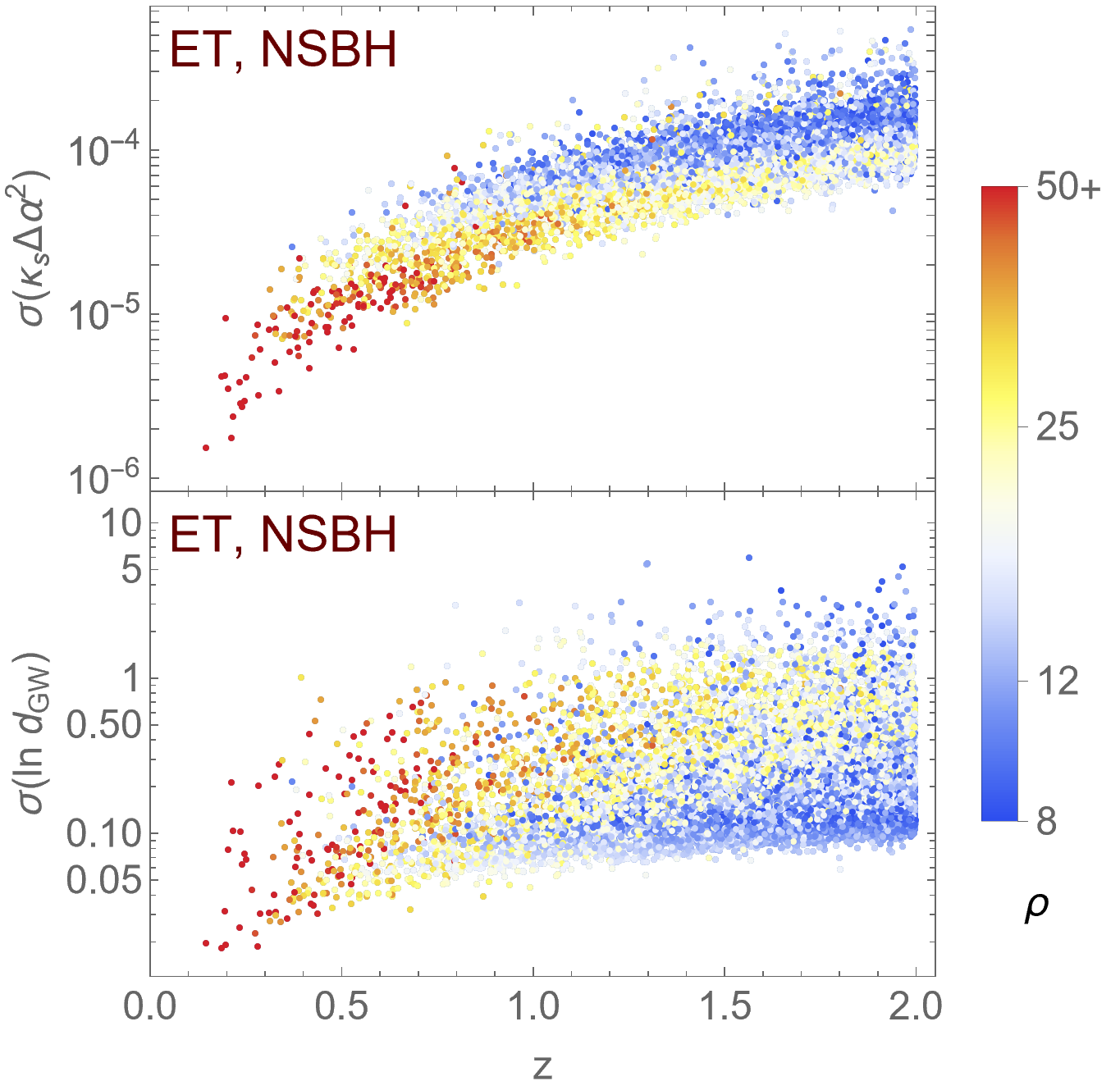}
    \vspace{0.1cm}
    \caption{Scatter plot of precision for the scalar charge $\kappa_s(\Delta{\alpha})^2$ and distances $d_{\rm GW}$ as a function of $z$ and color coded according to the logarithm of the S/N $\rho$ for the case of BNS and NSBH events. Each panel contains 10000 Monte Carlo simulations. Higher $\rho$ is correlated with better precision, but for a given $\rho$ the scatter is large. We assume flat $\Lambda$CDM to obtain $z$.
    }\label{fig:forecasts-scatter}
\end{figure}

Since parameter precision is highly correlated with $\rho$, we further explore this relation in Figure~\ref{fig:forecasts-scatter}. We plot as a function of redshift parameter precision for both $\kappa_s(\Delta{\alpha})^2$, the dominant phase parameter, and $\ln d_{\rm GW}$, the most interesting amplitude parameter. Each point is color-coded according to $\rho$, using a logarithm color scale. This figure makes it apparent how the higher significance events are clustered in the lower redshift intervals. It also shows how $\rho$ does not completely determine precision, as for a given $\rho$ precision can still vary by a factor of around 5. In particular, $d_{\rm GW}$ has a weaker correlation with $\rho$ than the $-1$PN term $\kappa_s(\Delta{\alpha})^2$. Both parameters have a stronger dependence on $z$ than on $\rho$.

One notices that the distance $d_{\rm GW}$ cannot be typically measured in high precision with single events. Even limiting to the events with $\rho\ge30$, the average error is above $30\%$ for ET. Only a fraction of events will have errors below $10\%$. For those events with EM counterpart the inclination may be constrained independently~\cite{Hotokezaka:2018dfi,Dhawan:2019phb,Alfradique:2022tox}, which improves moderately the distance precision. In principle, one can also improve the estimation by binning several events at similar redshifts, but then one needs to assume that the effective Planck mass $F(\phi_s)$ in Eq.~\eqref{dua} is the same for every merger, which can only be true if the screening effect is independent of the source, which is an extra assumption.

\section{Conclusions}
\label{conclusion}

In this paper, we worked out the inspiral waveform of GWs propagating on the cosmological background for most general Horndeski theories with a luminal GW propagation speed. The waveform contains a single parameter combination at $-1$PN level, denoted $\kappa_s(\Delta\alpha)^2$, which is expected to vanish for BH-BH mergers, and to be maximal for NSBH. In this case, $\kappa_s(\Delta\alpha)^2$ depends on Horndeski parameters and on the NS equation of state, and carries therefore important information on fundamental physics. For completeness, we also expanded the waveform to include the 0.5PN modified gravity term and higher PN terms in GR up to 1.5PN so as to study the impact on the estimation of the Horndeski parameter. In particular, we take into account the spin-orbit parameter which allows us to break the mass ratio degeneracy.

We forecast the measurement of $\kappa_s(\Delta\alpha)^2$ from aLIGO and ET by distributing sources in the sky with random masses, spins, distance, inclination, and polarization according to expected distributions and with signal-to-noise higher than 8. We evaluate for each event the constraints on our set of nine parameters with a Fisher Matrix approximation.
We therefore end up with obtaining a distribution of expected marginalized constraints on each parameter. Focusing on the Horndeski parameter, and quoting only the average precision  for Advanced LIGO, we see that one can expect to measure the new parameter $\kappa_s(\Delta\alpha)^2$ with $\sim$0.0004 for NSBH events with standard spin priors. For BNS, the constraints improve by one order of magnitude. ET would improve the constraints typically by a factor of five.

In both BD theories and the model of spontaneous scalarization of Damour and Esposito-Farese, we have $\kappa_s=1/2$ and hence the
limit $\kappa_s(\Delta\alpha)^2 \le 0.0004$ translates to
$|\Delta \alpha| \le 0.03$. On using the approximate relation (\ref{alQre}), it should be possible to give an upper bound on $|Q|$ at the level of order $0.01$. In the case of spontaneous scalarization, the scalar charge can be as large as $|\hat{\alpha}_A| \simeq 0.3$ for the maximum NS mass. Thus, the future NS events will offer the opportunity to exclude such highly scalarized NSs. A kinetic screening model of spontaneous scalarization induced by the term $\mu_2 X^2$ in $G_2$ can reduce $|\hat{\alpha}_A|$ down to the order 0.01, so it is also interesting to probe the compatibility of such a scenario with future GW observations. Thus, the scalar charge will be certainly a measurable quantity if present even by single events with current detectors. These bounds are complementary to those achievable with pulsar timing, which are currently much stronger but limited to few sources and to our own galaxy.

We conclude from the above that for testing Horndeski theories there is an interesting trade-off between detecting BNS and NSBH mergers. BNS should have a smaller $-1$PN signal (since this term is proportional to the difference in the scalar charge parameter $\alpha$ in each NS, leading to a partial cancellation) but much higher precision. BNS coalescences may also be more frequent in the Universe than NSBH, but NSBH can be detected at greater distances. In any case, both types of coalescence will allow very precise tests of GR in the coming years.

It turns out that in Horndeski theories, just as in GR, the phase and amplitude parameters in the waveform can be considered essentially independent. Likewise, although the Horndeski scalar charge affects both amplitude and phase, since the latter is much more constraining than the former, the effect of the new amplitude degree-of-freedom has negligible impact in the likelihood. Therefore, the inclusion and subsequent marginalization of the $-1$PN term does not affect the inferred distance precision. Similarly, the inclusion of the new 0.5PN phase term leads only to marginal, percent-level changes in the final precision.

\section*{Acknowledgements}

We thank Viviane Alfradique, Yurika Higashino, Atsushi Nishizawa, and Hiroki Takeda for useful discussions and comments. This study is financed in part by the Coordenação de Aperfeiçoamento de Pessoal de Nível Superior - Brasil (CAPES) - Finance Code 001. We acknowledge support from the CAPES-DAAD bilateral project  ``Data Analysis and Model Testing in the Era of Precision Cosmology''.  MQ is supported by the Brazilian research agencies CNPq, FAPERJ (grant~E-26/201.237/2022) and CAPES, and is thankful to University Heidelberg for hospitality and support. ST is supported by the Grant-in-Aid for Scientific Research Fund of the JSPS Nos.~19K03854 and 22K03642. LA acknowledges support from DFG project  456622116. LA thanks JSPS for support and the Waseda University for kind hospitality during the early phases of this project. RS acknowledges partial support by CNPq grant N.~310165/2021-0, and by FAPESP under grants N.~2021/14335-0 and 2022/06350-2.

\appendix

\section{Forecasts for minimal spin priors}\label{app:prior}

\begin{figure}[t]
    \centering
    \includegraphics[width=0.44\textwidth]{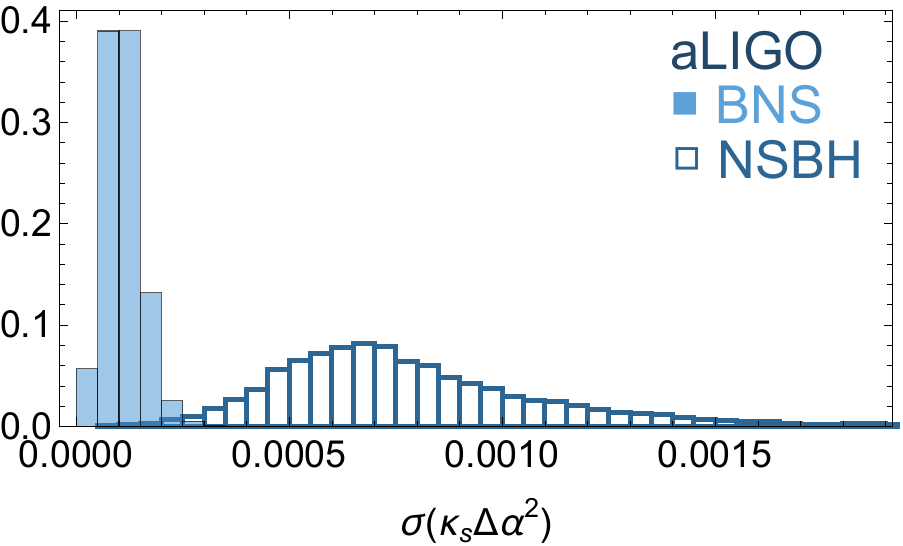}\quad
    \includegraphics[width=0.45\textwidth]{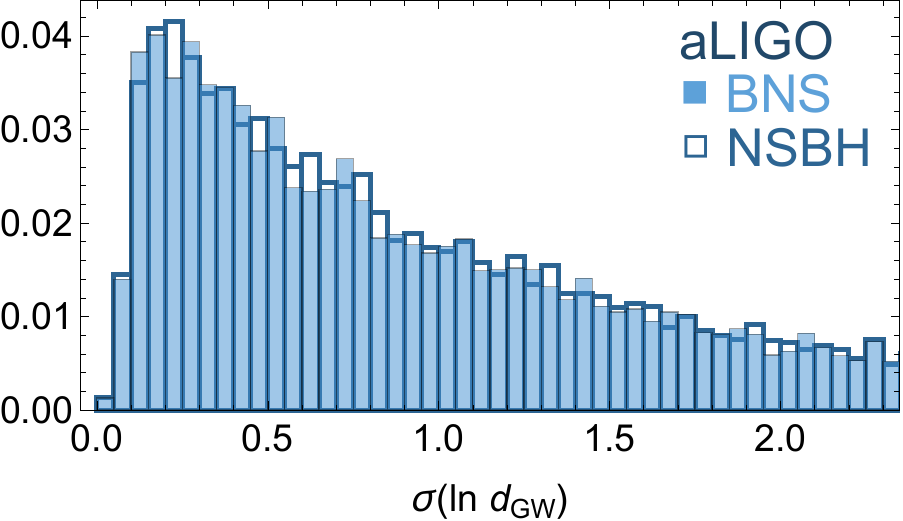} \\
    \vspace{0.3cm}
    \includegraphics[width=0.445\textwidth]{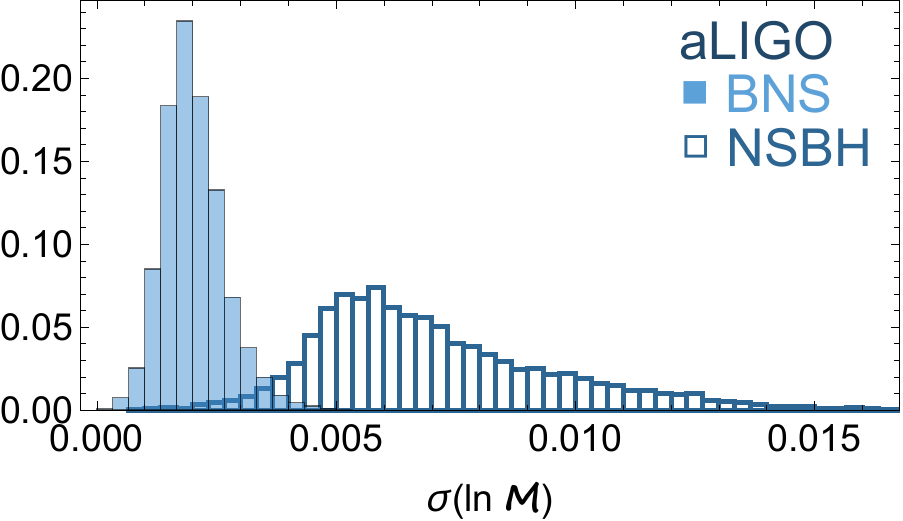}
    \quad
    \includegraphics[width=0.445\textwidth]{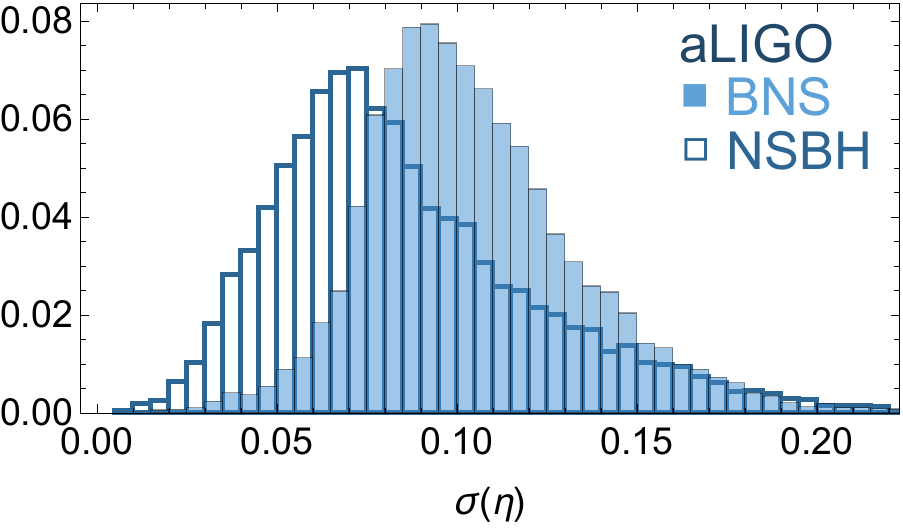} \\
    \vspace{0.3cm}
    \includegraphics[width=0.45\textwidth]{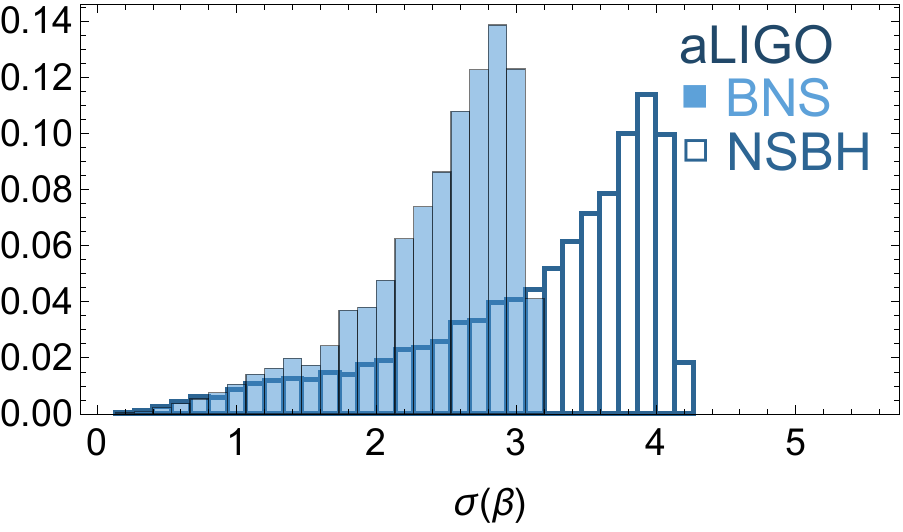}
    \caption{Same as Figures~\ref{fig:forecasts-alpha} and \ref{fig:forecasts-extra} but for the minimal spin priors: $\chi_{\rm NS}, \chi_{\rm BH} \in {\cal U}[-1, 1]$. We also include the forecasts for the spin-orbit parameter $\beta$. Such a generous spin prior results in a significant decrease in precision in all parameters except $d_{\rm GW}$.  }\label{fig:min-prior}
\end{figure}

Figure~\ref{fig:min-prior} illustrates how much the constraints degrade if we replace the spin priors by a minimalistic assumption that they simply obey the fundamental GR bounds: $\chi_{\rm NS}, \chi_{\rm BH} \in {\cal U}[-1, 1]$. As can be seen, the use of the minimal spin prior makes uncertainties in all parameters larger, except distance which is unaffected. The degradation is larger for BNS than for NSBH, because in the latter case the likelihood on $\beta$ is tighter, so the prior influence is diminished. In particular, for BNS (NSBH), we find that the precision loss is around: 2.4 (2.0) for $\kappa_s(\Delta \alpha)^2$, 4.4 (2.6) for $\ln{\cal M}$, 13 (3.3) for $\eta$ and 15 (2.9) for $\beta$. The large loss of precision for $\eta$ in BNS is a result of the large correlation between it and $\beta$.

\section{Estimating the scalar charge signal in BNS systems}\label{app:Delta-alpha}

Here we want to estimate how promising are BNS events compared to NSBH ones. Since the magnitude of the possible scalar charge depends not only on the particular Horndeski theory, the idea here is simply to estimate how much suppression there is on the quantity $(\Delta\alpha)^2\equiv (\alpha_A-\alpha_B)^2$ compared to either $\alpha_A^2$ and $\alpha_B^2$. For instance, if a NS $A$ has $\alpha_A = 0.1$, it will produce a GW signal four times stronger if its companion is $(i)$ a BH instead of $(ii)$ a NS with $\alpha_B = 0.05$. However, since we have shown that the typical uncertainties in the scalar charge are ten times smaller in BNS systems than for NSBH, in this case option $(ii)$ would still be easier to detect.

\begin{figure}[t]
    \centering
    \includegraphics[width=0.465\textwidth]{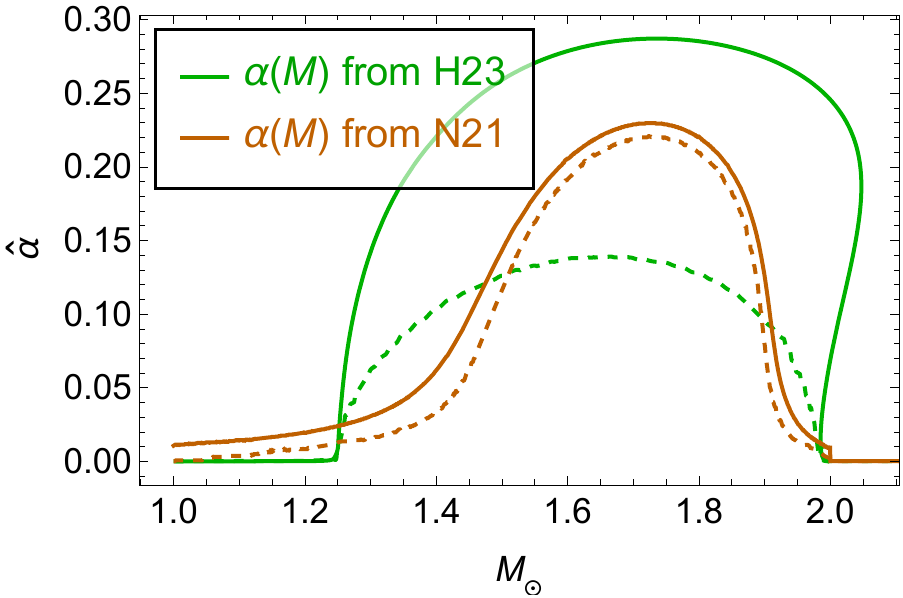}\qquad
    \includegraphics[width=0.43\textwidth]{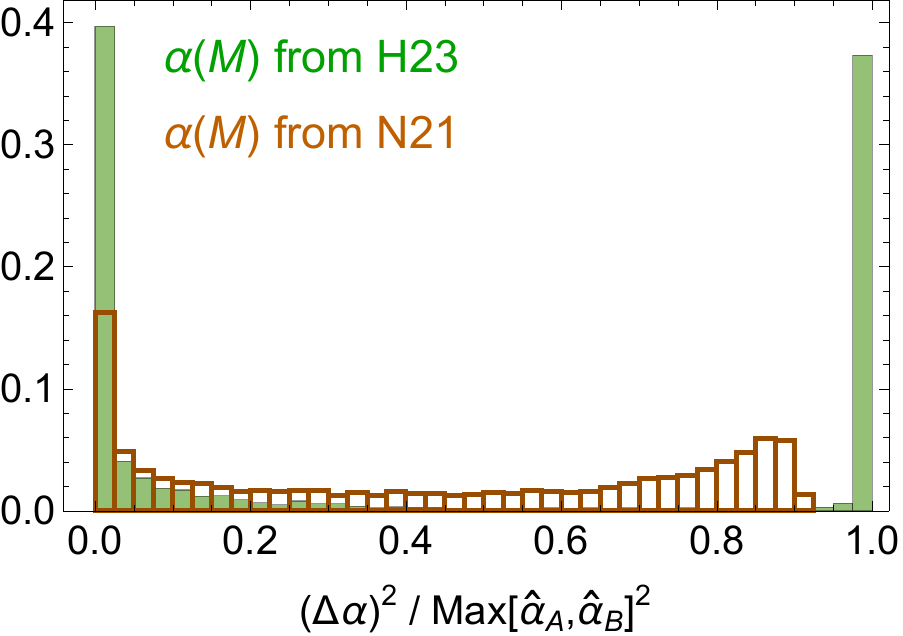}
    \caption{ \emph{Left:} A reproduction of the curves $\hat\alpha(M)$ from N21~\cite{Niu:2021nic} (top two orange curves of their Fig.~2) and H23~\cite{Higashino:2022izi} (top two curves of their Fig.~1b).
    \emph{Right:} 10000 BNS Monte Carlo samples of the ratio of $\Delta \alpha$ and the maximum $\hat\alpha$ in both NS using the solid curves on the left panel. Note that many BNS have $\Delta \alpha$ values which are a sizeable fraction of the maximum $\hat\alpha$.
    }\label{fig:Delta-alpha}
\end{figure}

To quantify this better, we use the estimates on how $\hat\alpha$ depends on the mass of a NS as computed both by~\cite{Niu:2021nic} (henceforth N21) and~\cite{Higashino:2022izi} (henceforth H23). Although there are differences, such as a phase transition around $M = 1.25 M_\odot$ in H23 and on the amplitudes of these curves, the overall shapes are qualitatively similar. So in Figure~\ref{fig:Delta-alpha} we depict the result of 10000 Monte Carlo samples of $(\Delta\alpha)^2$ divided by the maximum value between $\alpha_A^2$ and $\alpha_B^2$ assuming that the NS masses are distributed uniformly in the range $1-2 M_\odot$. This ratio should not depend strongly on the strength of the scalar charge of a given theory, and thus gives a good assessment of the how promising BNS are relative to NSBH for our purposes. We note that many BNS have $\Delta \alpha$ values which are a sizeable fraction of the maximum $\hat\alpha$. In particular, on average we have $(\Delta\alpha)^2$ = 44\% (43\%) of Max$(\hat\alpha_A^2,\hat\alpha_B^2)$ assuming  N21 (H23), using the solid curves of Figure~\ref{fig:Delta-alpha}. Using the dashed curves result in similar values, namely 56\% (48\%) for N21 (H23). Likewise, assuming a tighter mass distribution for the NS also does not change much these numbers.

We conclude that each BNS event is on average more promising than each NSBH event for detecting a possible scalar charge in GWs.

\bibliographystyle{mybibstyle}
\bibliography{bib}

\end{document}